\documentclass[%
 reprint,
 amsmath,amssymb,
 aps,
]{revtex4-2}

\usepackage[spanish,mexico,es-noindentfirst,british]{babel}
\usepackage{mathtools,autobreak,stmaryrd}
\usepackage[usenames,dvipsnames,svgnames,table]{xcolor}

\usepackage{tikz,scalefnt}
\usetikzlibrary{decorations.markings,positioning,shapes.misc,calc,patterns,angles,quotes}
\usetikzlibrary{external}

\usepackage{graphicx,multirow}
\usepackage[caption=false,font=Large]{subfig}
\usepackage{dcolumn}
\usepackage{bm}

\newcommand{\hatx}{\hat{x}}\newcommand{\hatp}{\hat{p}}
\newcommand{\vecx}{\mathbf{x}}
\newcommand{\vecE}{\mathbf{E}}

\DeclarePairedDelimiter\bra{\langle}{\rvert}				
\DeclarePairedDelimiter\ket{\lvert}{\rangle}				
\DeclarePairedDelimiterX\braket[2]{\langle}{\rangle}{#1 \delimsize\vert #2}			
\DeclarePairedDelimiterX\ketbra[2]{\vert}{\vert}{#1 \rangle\delimsize\langle #2}	
\DeclarePairedDelimiter\prom{\langle}{\rangle}				

\DeclarePairedDelimiter\Bparen{\Big(}{\Big)}
\DeclarePairedDelimiter\bparen{\big(}{\big)}

\DeclarePairedDelimiter\pen{\llbracket}{\rrbracket}				

\DeclarePairedDelimiterX\AConm[2]{\big\lbrace}{\big\rbrace}{#1 \,\text{\LARGE,}\, #2}		
\DeclarePairedDelimiterX\Conmu[2]{\big[}{\big]}{#1 \,\text{\LARGE,}\, #2}					

\DeclareMathOperator{\sgn}{sgn}
\DeclareMathOperator{\Si}{Si}

\begin{document}

\preprint{APS/123-QED}

\title{A Dynamical Quantum Daemon}

\author{J. I. Castro--Alatorre}
	\email[]{jcastroa@ifuap.buap.mx}
	\affiliation{%
	Instituto de Física, Benemérita Universidad Autónoma de Puebla, Apartado Postal J-48, 72570 Puebla, México
}%
\author{D. Condado}
	\email[]{dcondado@ifuap.buap.mx}
	\altaffiliation[Also at ]{Facultad de Ciencias Físico Matemáticas, Benemérita Universidad Autónoma de Puebla, 72570 Puebla, México}
\author{E. Sadurní}%
	\email[]{sadurni@ifuap.buap.mx}
	\affiliation{%
	Instituto de Física, Benemérita Universidad Autónoma de Puebla, Apartado Postal J-48, 72570 Puebla, México
}%


\date{\today}

\begin{abstract}
	We study the irreversibility {\it à la} Maxwell from a quantum point of view,
	involving an arbitrarily large ensemble of independent particles,
	with a daemonic potential that is capable of inducing asymmetries in the evolution,
	exhibiting new perspectives on how Maxwell's apparent paradox is posed and resolved dynamically.
	In addition, we design an electromagnetic cavity, to which dielectrics are added, fulfilling the function of a daemon.
	Thereby, this physical system is capable of cooling and ordering incident electromagnetic radiation.
	This setting can be generalized to many types of waves, without relying on the concept of measurement in quantum mechanics.
\end{abstract}

\maketitle

\section{\label{sec:level1}Introduction}
	Maxwell's daemon \cite{Maxwell1871Theory} is a theoretical construct that has played an important role in the history of physics,
	especially in thermodynamics and information theory.
	Such an entity has undoubtedly been witnessing scientists strive with its paradoxical behavior
	since its appearance in 1867 as part of a {\it Gedankenexperiment} discussing the limitations of the second law of thermodynamics \cite{Szilard1929, Brillouin1951, Landauer1961, Bennett1982, Lloyd1997, Maruyama2009, ScullyScully2010, KimSagawaUeda2011, Sagawa2011, Plesch2014, LeffRex2014, Lutz2015, Rex2017, Naghiloo2018}.
	Maxwell's daemon has not been emulated dynamically so far in the quantum realm;
	instead we have seen informational treatments which, in general, are based in measurements and feedback
	\cite{Scully2001,PriceBannerman2008,Camati2016,Cottet2017,Elouard2017} in various types of arrangements,
	such as photonic setups \cite{Ruschhaupt2006,Vidrighin2016}, ultracold atoms \cite{Kumar2018},
	superconducting quantum circuits \cite{Masuyama2018}, QED cavities \cite{NajeraSantos2020},
	quantum dots \cite{AnnbyAndersson2020} and electronic circuits \cite{Chida2017,Schaller2018}.
	
	This article proposes to study the irreversibility {\it à la} Maxwell from a dynamical point of view
	involving an arbitrarily large ensemble of independent particles.
	The general purpose of a dynamical realization with irreversibility,
	opens the door to other undulatory systems, either classical or quantum-mechanical,
	as it does not involve the collapse mechanism of the wave upon hypothetical measurements of velocity performed by a daemon.
	Instead, it should be possible to implement an interaction which depends on the momentum of the wave, i.e. its Fourier component.
	Thus, we expect that a point-like defect (such as potential barriers) which also incorporates non-local effects should
	reproduce reasonably the classical division into two compartments, plus interference effects that we shall discuss carefully.
	Hopefully, such treatment can be implemented in electromagnetic cavities and even elastic waves.	
	
	In section \ref{sec:CMM} there is a review about the phase space of a Maxwell's daemon in a classical setting,
	introducing a potential that emulates the daemon's action on the particles involved.
	Then, in section \ref{sec:QMG} there is a daemon's generalization in a quantum mechanical context.
	In addition, the formalism of Green's functions is introduced, obtaining thereby a new closed expression for these objects.
	Likewise, in section \ref{sec:EQD} an emulation of a quantum dynamical daemon in an electromagnetic billiard is presented.
	Finally, a design of an experimental setup using cavities is studied; the conducting cavity is a physical system that is
	capable of cooling and ordering incident electromagnetic radiation.
	
\section{A Classical Model for Maxwell's Daemon using Potentials}\label{sec:CMM}
\begin{figure}[b]\centering
		\includegraphics[width=0.45\textwidth]{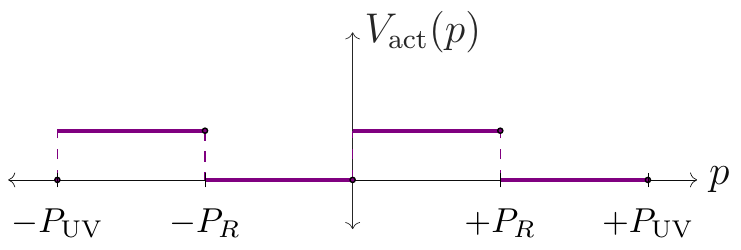}
		\caption{Activation potential $V_\text{act}(p)$ defined in \eqref{eq:VactSgn} with reference momentum $P_R$ and ultraviolet cutoff $P_\text{UV}$.}
		\label{fig:VactPUV}
	\end{figure}

	In the classical context, the particles are considered independent,
	reason why the action of the daemon is separable by particle until giving a Hamiltonian formulation of individual particle.
	So, in order to simulate a microscopic intelligent entity,
	that controls a door between two boxes to filter individual particles based on their energy,
	we define a Maxwell's daemon potential, acting on a particle whose Hamiltonian is
	\begin{equation}\label{eq:HamV}
		H= \frac{p^2}{2m} +V_\text{int}(x,p).
	\end{equation}
	$V_\text {int}$ represents the interaction between the daemon and a particle $V_0$ at $x =0$, i.e.
	\begin{equation}
		V_\text{int}(x,p) = V_0 \delta(x)V_\text{act}(p),
	\end{equation}
	where $V_\text{act}(p)$ is an activation function that allows the daemon to open
	the shutter	between compartments according to the particle's momentum.
	Indeed, we have the following expression
	\begin{equation}\label{eq:VactSgn}\begin{split}
		V_\text{act}(p) &= f_-\bparen{|p|} \sgn(p) + f_+\bparen{|p|}
	\end{split}\end{equation}
	with
	\begin{align}\begin{autobreak}
		2 f_\pm\bparen{|p|} = \Theta\Bparen{P_R-|p|} \pm \Theta\Bparen{|p|-P_R} \mp \Theta\Bparen{|p|-P_\text{UV}}
	\end{autobreak}\end{align}
	where $P_R$ is a reference momentum that will discern which particles will be affected by the potential,
	depending on the momentum, while the particle's position is captured by the presence of $\delta(x)$.
	In addition, $P_\text{UV}$ is an ultraviolet cutoff when necessary. (See fig.~\ref{fig:VactPUV}.)
	
\begin{figure}[h]\centering
		\includegraphics[width=0.45\textwidth]{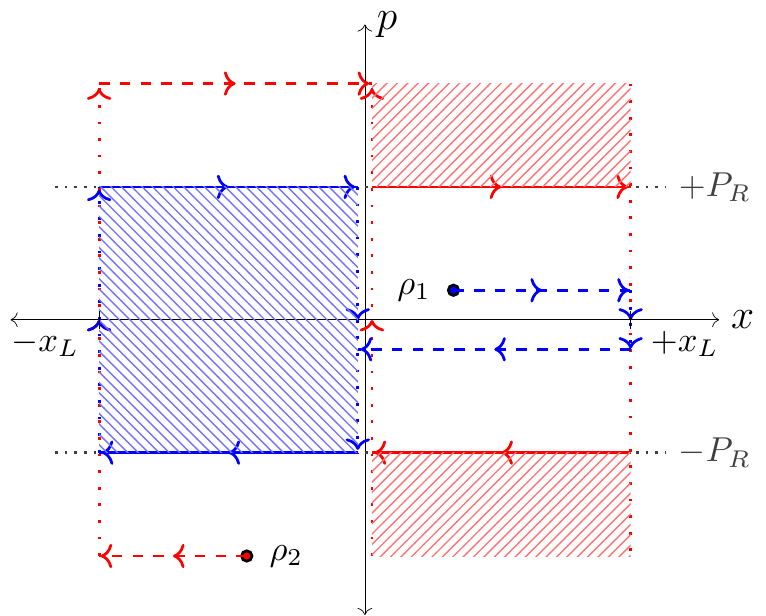}
		\caption{Phase space evolution of two particle ensembles $\rho_1(|p|<|P_R|)$ \& $\rho_2(|p|>|P_R|)$.
		The activation potential separates the particles according to the reference momentum, leaving two well differentiated zones.}
		\label{fig:EFEnsamble}
	\end{figure}

	To appreciate the effect of this potential suppose two ensembles, $\rho_1$ and $\rho_2$, shown in fig.~\ref{fig:EFEnsamble}:
	$\rho_1$ represents a collection of independent particles in the first quadrant with a right directed momentum less than $P_R$,
	therefore,	when the system evolves, the phase space corresponding to the zone $ -x_L <x <0 $ and $ | p | <| P_R | $ will be filled.
	Conversely, $ \rho_2 $ represents a collection of independent particles in the third quadrant with a left directed momentum greater than $ P_R $,
	so,	when the system evolves, the phase space corresponding to the zone $ 0 <x <x_L $ and $ | p |> | P_R | $ will be filled.
	It should also be noted that, after the selection process has taken place,
	the system reaches an equilibrium such that each compartment possesses temperatures $T$ such that
	\begin{align}
		T_2\Bparen{x>0} > T_1\Bparen{x<0}.
	\end{align}
	Therefore $V_\text{act}(p)$ effectively separates the particles into two well differentiated zones according to their momentum.	
	
\section{Quantum Mechanical Generalization}\label{sec:QMG}
	We shall preserve the hermiticity of the Hamiltonian in the Schrödinger equation,
	so we consider a potential $ \hat{V}_\text{int}(\hatx, \hatp) $ as an operator with the property
	\begin{align}\label{eq:PotHer}\begin{autobreak}
		\hat V(\hatx,\hatp) = \frac{V_0}{2}
		\Big( V_\text{act}(\hatp)\delta(\hatx) + \delta(\hatx)V_\text{act}(\hatp) \Big)
	\end{autobreak}\end{align}
	which is manifestly hermitian because of $(1/2)(V+V^\dagger)$.
	This renders the Schrödinger equation as
	\begin{align}\label{eq:DdMQMEq}\begin{autobreak}
		-\frac{\hbar^2}{2m}\nabla^2\Psi(x,t)
		+ \frac{V_\text{int}(x,-i\hbar\nabla)
		+ V_\text{int}^\dagger(x,-i\hbar\nabla)}{2}\Psi(x,t)
		= i\hbar\partial_t\Psi(x,t).
	\end{autobreak}\end{align}
	The goal is to solve \eqref{eq:DdMQMEq} using the Green's functions as shown in section \ref{sec:EFGf}.
	(See appendix \ref{ap:a} for a detailed procedure.)

	\subsection{Exact Form of Green's Function}	\label{sec:EFGf}
		The analysis that will be implemented in this work is based on the use of a $ G_\text{Daemon} $ function that solves the following problem
		\begin{align}\label{eq:GreenDem}
			\Bparen{H+V(x,p)-E}G_\text{Daemon}= \delta(x-x'),
		\end{align}
		where $V(x,p)$ is given by the daemonic interaction in \eqref{eq:DdMQMEq}.
		The solution techniques for this Green's function consist of
		i) {\it Analytical method:} Analytical evaluation of the Born--Oppenheimer series in all its terms,
			where energy dependent coefficients are thrown at all possible orders,
		ii) {\it Numerical methods:} Spectral decomposition through the eigensystem of interest,
		iii) {\it Numerical simulation of the dynamics:} Given the energy-dependent Green's function,
			we analyze the pole structure of that expression to recover the time-dependent harmonic dependence through
			the Fourier's semi-transform of a known meromorphic function.
		Following the essence of \cite{MoshinskySadurni2007},
		it would be possible to find the Green's function for the problem in question in an analytical way.
		This procedure is explained in detail in appendix \ref{ap:a},
		and it constitutes a new result in the computation of Green's functions for non-local delta perturbations,
		not found in the standard literature \cite{GroscheSteiner1998}.
		The result is
		\begin{subequations}\label{eq:GreenPotencialDelta}
		\begin{align}\begin{split}
		G_p(x, &x', E) = G_0(x,x',E) \\
			&+ \frac{G_0(x,0,E)G_0(0,x',E)Q_3(E)}{1+P_2(0,E)-G_0(0,0,E)Q_3(E)} \\
			&-\frac{G_0(x,0,E)R_1(x',E)\Bparen{1+P_2(0,E)} }{1+P_2(0,E)-G_0(0,0,E)Q_3(E)} \\
			&-\frac{P_2(x,E)\Bparen{G_0(0,x',E)-G_0(0,0,E)R_1(x',E)} }{1+P_2(0,E)-G_0(0,0,E)Q_3(E)}
		\end{split}\end{align}
		where
		\begin{equation}
			P_1(x',E) = \int dx\,\widetilde V(x)G_0(x,x',E),
		\end{equation}
		\begin{equation}
			P_2(x,E) = \int dy\,G_0(x,y,E)\widetilde V(-y),
		\end{equation}
		\begin{equation}
			Q_1(E) = \int dx\,\int dy\,\widetilde V(x)G_0(x,y,E)\widetilde V(-y),
		\end{equation}
		\begin{equation}
			Q_2(E) = \int dx\,\widetilde V(x)G_0(x,0,E),
		\end{equation}
		\begin{align}
			R_1(x',E) = \frac{P_1(x',E)}{1+Q_2(E)}, && Q_3(E) = \frac{Q_1(E)}{1+Q_2(E)},
		\end{align}
		and $\widetilde V(y)$ is the Fourier's transform of the activation potential.
		\end{subequations}
		Subsequently, employing the free Green's function in a container $G_0^C$, i.e.
		\begin{align}\begin{autobreak}
			G_0^\text{C}(x,x',E) =
			\frac{2}{L}\sum_{m=1}\frac{\sin(\kappa_{2m}x)\sin(\kappa_{2m}x')}{E_{2m}-E}
				+ \frac{2}{L}\sum_{m=1}\frac{\cos(\kappa_{2m-1}x)\cos(\kappa_{2m-1}x')}{E_{2m-1}-E}
		\end{autobreak}\end{align}
		with $\kappa_n=n\pi/L$ and eigenenergies $E_n = \frac{1}{2}\hbar^2\kappa_n^2$,
		the Green's function in \eqref{eq:GreenPotencialDelta} but now in a container is
		\begin{align}\begin{split}
			&G_p^\text{C}(x,x',E) = \quad G_0^\text{C}(x,x',E) \\
			&+\frac{P_1^\text{C}(x,E)G_0^\text{C}(0,x',E)-G_0^\text{C}(x,0,E)P_1^\text{C}(x',E)}{1-G_0^\text{C}(0,0,E)Q_1^\text{C}(E)} \\
			&+\frac{G_0^\text{C}(x,0,E)G_0^\text{C}(0,x',E)Q_1^\text{C}(E)}{1-G_0^\text{C}(0,0,E)Q_1^\text{C}(E)} \\
			&-\frac{P_1^\text{C}(x,E)G_0^\text{C}(0,0,E)P_1^\text{C}(x',E)}{1-G_0^\text{C}(0,0,E)Q_1^\text{C}(E)},
		\end{split}\end{align}
		where $P_1^\text{C}$ and $Q_1^\text{C}$ are the integrals evaluated with $G_0^C$.
		From the previous expression the identification of the symmetric and antisymmetric part of the Green's function is possible.
		The only terms that contribute to the antisymmetric part come from the daemonic perturbation,
		as the first and third terms are manifestly symmetric.
	
	\subsection{Pole Structure Analysis}
		The integrals in \eqref{eq:GreenPotencialDelta} can be done in terms of sine-integral functions (Si) \cite{Abramowitz1965}
		using in addition the Fourier's transform of the potential described in fig.~\ref{fig:VactPUV}, e.g.
		\begin{subequations}
		\begin{equation}
			\widetilde V(\pm y) = \frac{\pm 1}{2i\pi y}\Bparen{1-2\cos P_R y},
		\end{equation}
		\begin{align}\begin{split}
			&P_1^\text{C}(x',E) = - P_2^\text{C}(x',E) = \\ &-\frac{2}{i\pi L}
			\sum_{n=1}\frac{\Si\bparen{\xi_+}-\Si\bparen{\xi_-}-\Si(n\pi)}{E_{2n}-E}\sin(\kappa_{2n}x'),
		\end{split}\end{align}
		\begin{equation}
			Q_1^\text{C}(E) =
			\frac{2}{\pi^2L}\sum_{n=1} \frac{\Bparen{\Si\bparen{\xi_+}-\Si\bparen{\xi_-}-\Si(n\pi) }^2}{E_{2n}-E},
		\end{equation}
		\begin{equation}
			Q_2^\text{C}(E) = 0
		\end{equation}
		with
		\begin{equation}
			\xi_\pm = \bparen{P_R\pm\kappa_{2n}}\frac{L}{2}.
		\end{equation}
		\end{subequations}
		Given the behavior of the $ \Si $ function in $ P_1^\text{C}$ and $Q_1^\text{C}(E)$, the following approximation can be made.
		The argument is written as
		\begin{align}\begin{split}
			&\Si\bparen{n\pi+a} - \Si\bparen{n\pi-a} \simeq \\
			&\frac{\pi}{2}-\frac{\pi}{2}\Theta\bparen{n-\pen{a/\pi}}+\frac{\pi}{2}\Theta\bparen{\pen{a/\pi}-n}+\pi\epsilon\,\delta_{n,\pen{a/\pi}}.
		\end{split}\end{align}
		where $a= P_R L/2$, $ \pen {a/\pi} $ represents the integer part and
		$\epsilon$ the fractional part of $a/\pi$
		and the step function is zero for the case $n=\pen{a/\pi}$.
		(The full procedure is in appendix \ref{ap:b}.)
		So the integrals are now approximated by
		\begin{subequations}
		\begin{align}\begin{autobreak}
			P_1^\text{C}(x,E)\simeq -\frac{2}{iL}
			\left(\sum_{n=1}^{\pen{a/\pi}-1}\frac{\sin(\kappa_{2n}x)}{E_{2n}-E}\right. 
			- \frac{1}{2}\sum_{n=1}^\infty\frac{\sin(\kappa_{2n}x)}{E_{2n}-E}
			+ \left.\frac{1}{2}\frac{(1+2\epsilon)}{E_{2\pen{a/\pi}}-E}\sin(\kappa_{2\pen{a/\pi}}x)\right)
		\end{autobreak}\end{align}
		and
		\begin{align}
			Q_1^\text{C}(E)\simeq\frac{1}{2L}\left(\sum_{n=1}^\infty\frac{1}{E_{2n}-E} - \frac{1}{E_{2\pen{a/\pi}}-E}\right).
		\end{align}
		\end{subequations}
		
		The purpose of this approach is to highlight the terms that depend strongly on the reference energy.
		And in this way, illuminate the presence of new poles created in the Green's function.
		
	\subsection{Numerical Dynamics}
		We proceed to discretize the Hamiltonian on a lattice.
		This will enable us to treat the problem as a matrix representation in a basis of point-like functions.
		Since the treatment is equivalent to a tight binding model in a crystal,
		it is advisable to use the first Brillouin zone to calculate the energies.
		In this way, the activation potential in \eqref{eq:VactSgn} will be non-zero in the intervals $[-\kappa_D,-\kappa_R]$ and $[0,\kappa_R]$,
		resulting in the action zones of the daemon according to the reference momentum $P_R\leftrightarrow\kappa_R$.
		This can be seen in the graph below in fig.~\ref{fig:EigenEn}.
		However, it is necessary to work in the quasi-parabolic energy regime that is below the {\it Dirac point} \cite{Yael2016} $(\varpi_\text{D})$,
		obtaining the upper graph, a parable with regions where the daemon acts.
		It is worth mentioning that the region of the potential for $p<-P_R$ is not bounded above as in the graph below.
		\begin{figure}[b]\centering
		\includegraphics[width=0.48\textwidth]{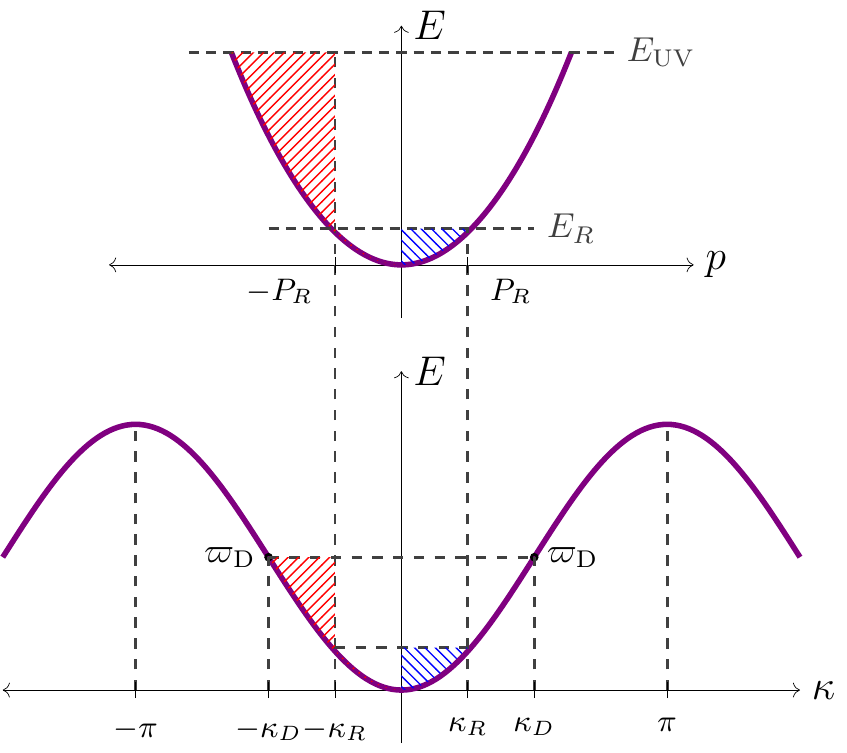}
		\caption{The graph below shows the energies in a tight binding model in a crystal,
		the coloured zones represent the activation potential in \eqref{eq:VactSgn} with reference momentum $P_R\leftrightarrow\kappa_R$.
		Also, using the quasi-parabolic energy regime below the {\it Dirac point} $(\varpi_\text{D})$, the upper graph is obtained.
		}
		\label{fig:EigenEn}
		\end{figure}%
		Therefore, the Hamiltonian's action on a plane wave
		\begin{equation}
		\ket{\kappa} = \frac{1}{\sqrt{2\pi}}\sum_{n} e^{i\kappa n}\ket{n},
		\end{equation}
		is no longer restricted to the unperturbed part plus the defect at the origin,
		instead we have a non-local effect that can be obtained directly by calculating the matrix elements at site $n$
		\begin{align}\begin{split}
		\bra{n}H\ket{\kappa} &= \frac{\hbar^2}{m a^2}\Bparen{1-\cos\kappa}\frac{e^{ikn}}{\sqrt{2\pi}} \\
			&+ \frac{1}{2i\pi n \sqrt{2\pi}}\Bparen{2\cos\bparen{\kappa_R n}-1-e^{-i\kappa_D n}} \\
			&- \sum_{n'}\frac{\delta_{n,0}}{2i\pi n'}\cdot\frac{e^{i\kappa n'}}{\sqrt{2\pi}}
				\Bparen{2\cos\bparen{\kappa_R n'}-1-e^{i\kappa_D n'}}.
 		\end{split}\end{align}	
 		where $a$ is the scale parameter.	
		Also, the Hamiltonian will be diagonalized using a discretized basis $\ket{n}$, as many sites as frequencies are necessary, i.e.
		\begin{align}\begin{split}\label{eq:nHnp}
		\bra{n}H\ket{n'} = &- \frac{\hbar^2}{2m a^2}\bparen{\delta_{n-1,n'} - 2\delta_{n,n'} + \delta_{n+1,n'} } \\
			&+ \frac{V_0}{2}\cdot\frac{\delta_{n',0}}{2i\pi n}\Bparen{2\cos\bparen{\kappa_R n}-1-e^{-i\kappa_D n}} \\
			&- \frac{V_0}{2}\cdot\frac{\delta_{n,0}}{2i\pi n'}\Bparen{2\cos\bparen{\kappa_R n'}-1-e^{i\kappa_D n'}}.
		\end{split}\end{align}
		Note that for the central element (the evaluation of the corresponding integrals at $n=0=n'$),
		\begin{align}
			\bra{0}H\ket{0} = \frac{\hbar^2}{2m a^2}\cdot 2 + \frac{V_0}{2}\cdot\frac{\kappa_D}{\pi}.
		\end{align}
		This shows that the potential at the location of the daemon is finite in a discretized setting, 
		and its intensity $V_0$ can be adjusted at will.
		
		Finally, the wave function $\Psi(t)$ at site $n$ is
		\begin{subequations}
		\begin{equation}
			\braket{n}{\Psi,t} = \sum_m \exp\bparen{-itE_m/\hbar}\braket{n}{m,E_m} \braket{m,E_m}{\Psi_0}
		\end{equation}
		where $E_m$ are the eigenvalues of the problem,
		$\braket{n}{m,E_m}$ are stationary functions i.e. eigenvectors,
		while $\braket{m,E_m}{\Psi_0}$ is the overlap (integral) of the initial condition with the basis, i.e.
		\begin{equation}
		\braket{m,E_m}{\Psi_0} = \sum_{n'} \braket{m,E_m}{n'} \braket{n'}{\Psi_0}.
		\end{equation}
		where $\braket{n'}{\Psi_0}$ is the initial condition.
		\end{subequations}
		
		\subsection{Entropy}
			The entropy is fundamental in the resolution of the apparent paradox,
			since a dynamic effect of apparent ordering is sought,
			rather than a quantum collapse and measurement effect.
			Since the von--Neumann equation does not capture the irreversibility phenomenon,
			a notion of entropy that captures disorder with respect to a specific basis (e.g. energy) is required.
			Shannon's definition of entropy is
			\begin{equation}\label{eq:ShEntropia}
			\sigma_\text{\tiny Sh} = - \sum_m \varrho_m \log \varrho_m.
			\end{equation}
			where the probabilities $\varrho_m$ will be given by the overlap (integral) of the wave function with the basis of the free problem, i.e.
			\begin{equation}
				\varrho_m = \big|\braket{m,E_m^\text{(0)}}{\Psi,t}\big|^2 =
				\left|\sum_{n'}\braket{m,E_m^\text{(0)}}{n'}\braket{n'}{\Psi,t}\right|^2.
			\end{equation}
			Likewise, the total entropy of the system must be estimated separately,
			as Shannon's entropy applies only to the particles inside the apparatus,
			but not the apparatus itself (which is not dynamically involved).
			To this end, we estimate the work done by the device on the trapped wave and {\it vice versa}.
			Also, using the principle of extensivity one can find a lower bound for the total entropy of the system $S_\text{t}$
			as the linear combination of the particle's entropy $S_\text{p}$ plus daemon's entropy $S_\text{d}$, i.e.
			\begin{subequations}
			\begin{equation}
				\Delta S_\text{t} = \Delta S_\text{p} + \Delta S_\text{d} \overset{?}{\geq} 0
			\end{equation}
			with $\Delta S_\text{p}\leq 0$ as the entropy of the particles must decrease because of daemonic action.
			Furthermore, one can find a lower bound for the daemon's entropy change in terms of the work done by the potential as
			\begin{equation}
				\Delta S_\text{d} = \int\frac{\delta'Q}{T} \;\sim\;
				\frac{\Delta Q}{\prom T} \geq \frac{1}{\prom T}\Delta V = \text{(const.)}\Delta V.
			\end{equation}
			From this, it follows that
			\begin{equation}
				\Delta S_\text{t} \geq \Delta S_\text{p} + \text{(const.)}\Delta V.
			\end{equation}
			\end{subequations}
			Therefore, when taking into account the work done by the daemon,
			its contribution must compensate for the partial entropy reduction.
			Indeed, for a box with two separated compartments with volume $\upsilon$, the change in the particle's entropy is
			\begin{subequations}
			\begin{equation}
				\Delta S_\text{p} = -\left(\frac{P_D}{T_D} + \frac{P_S}{T_S}\right) \upsilon\log 2,
			\end{equation}
			where $P_{D,S}$ and $T_{D,S}$ are the pressure and temperature for the right (D) and left (S) compartment, computed as ideal gases.
			Moreover, the internal energy $U\propto P\upsilon$, so
			\begin{equation}
				\Delta S_\text{p} \propto - \beta_\text{B}\left( U_D + U_S \right)\log 2,
			\end{equation}
			\end{subequations}
			where $\beta_\text{B}$ is the thermodynamic beta and $U_{D,S}$ the corresponding lateral internal energy.
	
		\subsection{Numerical Results}
			The Shannon analogue of a Boltzmann thermal distribution
			(e.g as understood by superposition of particle's number in photonic states) \cite{Schleich2001}
			can be used as an appropriate initial condition for box states.
			The idea is to monitor the evolution of such a state and its subsequent ordering. We have:
			\begin{equation}
			\psi_0^\text{B}(\beta,n)= \sum_{q=1}^{N_\text{max}} \exp\Bparen{-\beta\bparen{q^2-1}} \sin\frac{q\bparen{n+N}\pi}{2N},
			\end{equation}
			here $ n $ is the site such that $-N\leq n\leq N$ (the daemon is at $n = 0$),
			$N_\text{max}=2N+1$ is the maximum number of $q$ box states that are meaningful in a discretized system,
			and $\beta $ is an order parameter which would correspond to
			\begin{equation}
				\beta = \frac{E_0}{k_\text{B}T}, \quad\text{with}\quad E_0 = \frac{\pi^2\hbar^2}{2mL^2}
			\end{equation}
			in thermodynamics.
			This probability overlaps with the components of the eigenvectors $\nu_m^{(n)}$ of \eqref{eq:nHnp}
			\begin{equation}
			\Psi_0^\text{B}(\beta,m) = \sum_{n=1}^{2N+1} \psi_0^\text{B}(\beta,n) {\nu_m^{(n)}}^*
			\end{equation}
			obtaining the wave function at the rescaled time $\tau$ ($= \hbar t/2ma^2$ [adim])
			\begin{equation}\label{eq:PsiBoltz}
			\Psi_B(\beta,n,\tau) = \sum_{m=1}^{2N+1}\exp\Bparen{-i \tau\,\Xi_m} \Psi_0^\text{B}(\beta,m) \nu_m^{(n)},
			\end{equation}
			where $\Xi$ ($= 2ma^2E/\hbar^2$ [adim]) are the rescaled eigenenergies of \eqref{eq:nHnp}.
						
			\begin{figure}[htpb]\centering
				\includegraphics[width=0.48\textwidth]{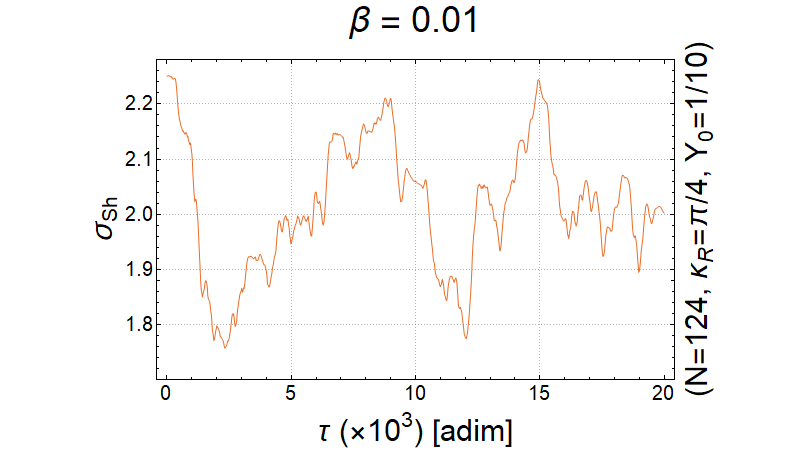}
					\caption{Entropy with $\beta=1/100$.
					A decrease of the entropy is seen at $\tau(\times10^{-3}) \sim$ 2 and 12.}
				\label{fig:EntropyE-SG124-BI4-VG10.png}
			\end{figure}%
			\begin{figure}[htpb]\centering
				\includegraphics[width=0.48\textwidth]{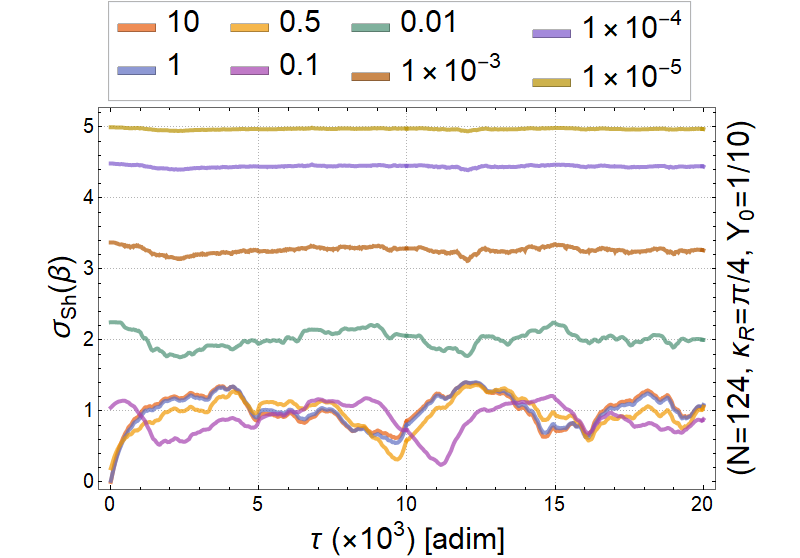}
					\caption{Comparative plot of entropies by varying the temperature value.}
				\label{fig:Entropien-SG124-BI4-VG10.png}
			\end{figure}%
			\begin{figure}[htpb]\centering
				\includegraphics[width=0.48\textwidth]{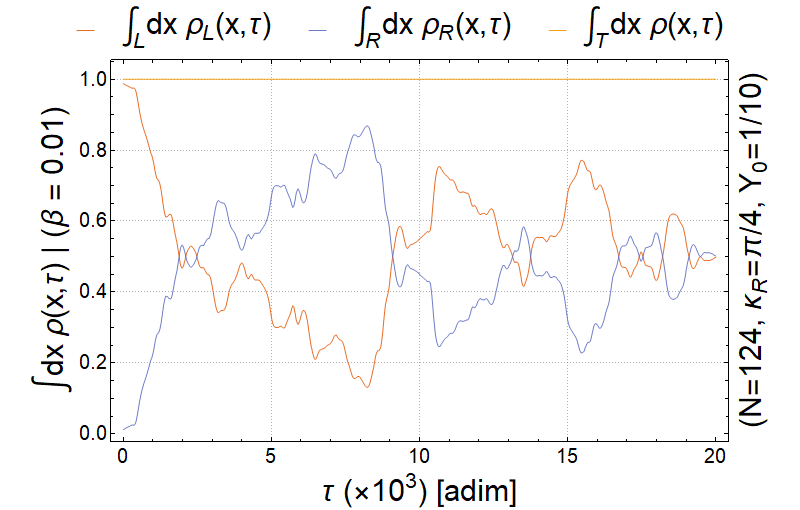}
					\caption{Lateral probabilities for the left (L, red line)
					and right (R, blue line) part of the container with $\beta=1/100$.}
				\label{fig:ProbE-SG124-BI4-VG10.png}
			\end{figure}%
			\begin{figure}[htpb]\centering
				\includegraphics[width=0.48\textwidth]{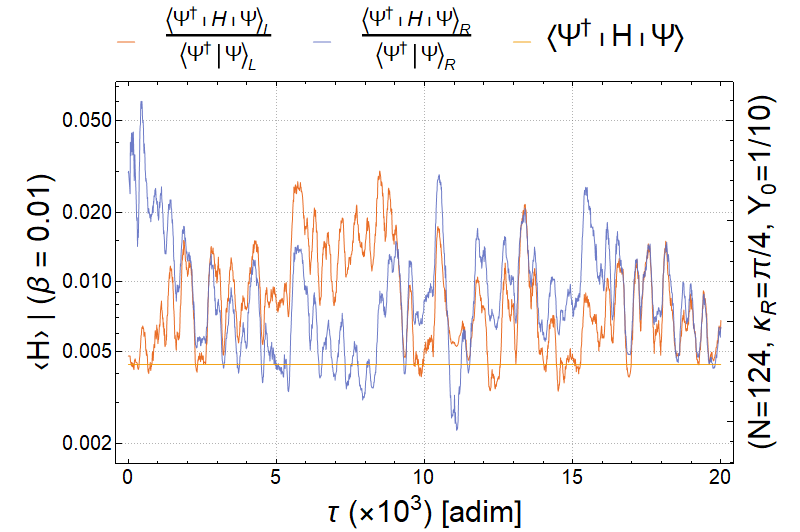}
					\caption{Average internal energy for the left (L, red line)
					and right (R, blue line) part of the container with $\beta=1/100$.}
				\label{fig:PsiHPsiE-SG124-BI4-VG10.png}
			\end{figure}%
			\begin{figure}[htpb]\centering
				\includegraphics[width=0.48\textwidth]{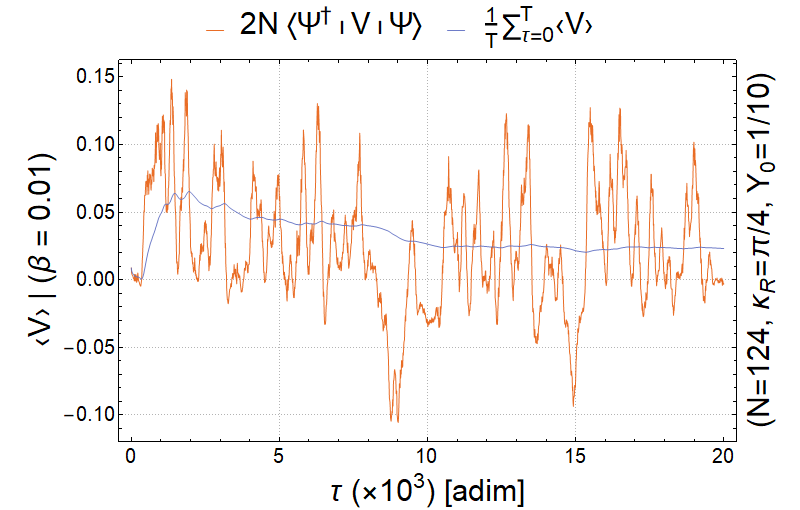}
					\caption{Average Potential Energy. Contributes significantly to the energy balance.
					The blue line indicates the time average at time $\tau$.
					Negative values implies work done by the wave on the daemon.}
				\label{fig:PsiVPsiE-SG124-BI4-VG10.png}
			\end{figure}%
			\begin{figure}[htpb]\centering
				\includegraphics[width=0.4\textwidth]{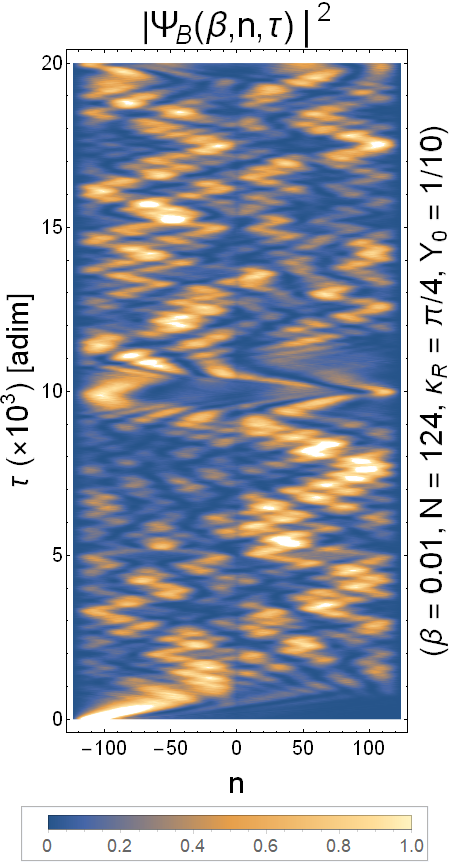}
					\caption{Evolution of a Boltzmann distributed wave packet in the interval
					$-124\leq n\leq 124$ (horizontal axis) at $\tau = 0$ (vertical axis),
					interacting with a Maxwell daemon located at $ n = 0$.
					The colouration exhibits the probability density,
					showing that for $5<\tau (\times10^{-3})< 10$ the wave packet is predominantly on the right side.}
				\label{fig:wBoltzPlotE-SG124-BI4-VG10.png}
			\end{figure}%
			\begin{figure}[htpb]\centering
				\includegraphics[width=0.48\textwidth]{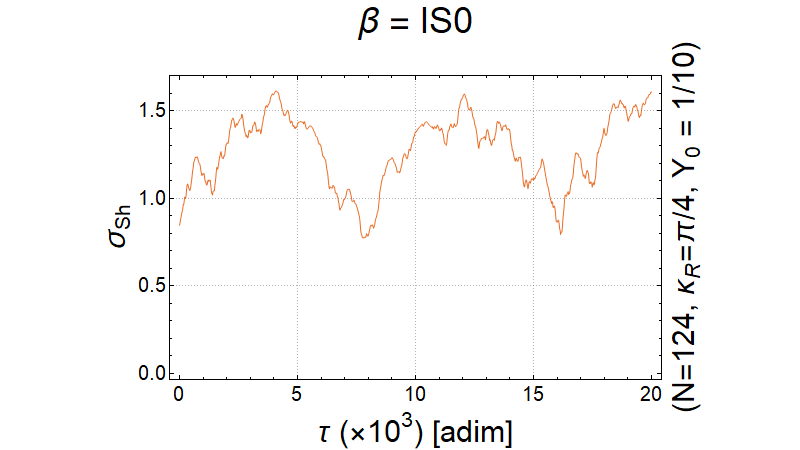}
					\caption{Entropy of a isoespectral wave packet.
					A decrease of the entropy is seen at $\tau(\times10^{-3}) \sim$ 8 and 16.}
				\label{fig:EntropyIS0-SG124-BI4-VG10.png}
			\end{figure}%
			\begin{figure}[htpb]\centering
				\includegraphics[width=0.4\textwidth]{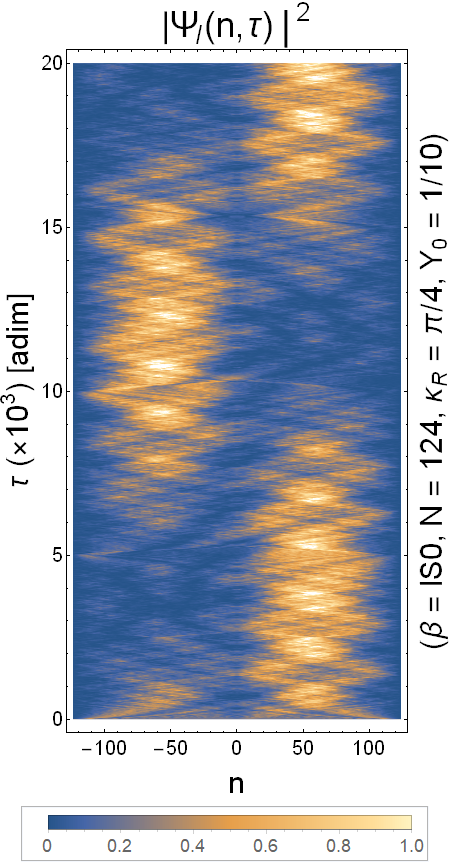}
					\caption{Evolution of a isoespectral wave packet in the interval
					$-124\leq n\leq 124$ (horizontal axis) at $\tau = 0$ (vertical axis),
					interacting with a Maxwell daemon located at $ n = 0$.}
				\label{fig:wBoltzPlotIS0-SG124-BI4-VG10.png}
			\end{figure}%
			
			An example of evolution is shown in fig.~\ref{fig:EntropyE-SG124-BI4-VG10.png}.
			The system size is $ 2N+1 = 249 $ sites, with scale parameter $a=L/2N$,
			the rescaled potential intensity is $\Upsilon_0 =\frac{1}{10}$ (=$ma^2V_0/\hbar^2$ [adim]),
			the reference momentum is $ \kappa_R = \pi/4 $ and $\beta =1/100$.		
			For relatively short times $\tau (\times10^{-3})\simeq 2$, a decrease of the entropy in \eqref{eq:ShEntropia} is appreciated.
			Then, between $\tau (\times10^{-3})\simeq 3$ to 9 the entropy increases which is explained by the natural wave expansion in each compartment,
			to decrease again at $\tau (\times10^{-3})\simeq 12$.

			Now we turn our attention to fig.~\ref{fig:Entropien-SG124-BI4-VG10.png}
			where we show a comparative plot of entropies by varying the temperature value.
			It is found that values $ 1/2 > \beta > 1/200 $ produce significant fluctuations for a potential intensity $\Upsilon_0 =\frac{1}{10}$.
			It should be stressed that for higher values of $\Upsilon_0$, the overall behavior shifts to larger values of beta.
			For very high temperatures, a highly disordered system in the energy basis has a tendency to fluctuate around its original entropic value
			(quasi-stationary behaviors), this implies that the effect is not strong in these cases.
			We have found, through these numerical results, that the role played by $\kappa_R$
			is partially decisive in the creation of box asymmetries in the evolution,
			as the intensity $\Upsilon_0$ is also important for small values of beta.
			However, we must stress that $\Upsilon_0$ cannot be taken as infinite, since all waves become trapped in such a case.
		
			In fig.~\ref{fig:ProbE-SG124-BI4-VG10.png} we can see asymmetries induced as time elapses,
			with must drastic effects occurring around $\tau (\times10^{-3})\simeq 8.5$
			where the difference between left and right probabilities (occupation) is large.
			Note that the effect is recurrent for larger times.
			In addition, the entropy has a minimum when the probability has a maximal rate of change with respect to time,
			implying that the daemon operates reaching a quasi-stationary regime,
			where there is no exchange of densities but there is entropic rise.

			In fig.~\ref{fig:PsiHPsiE-SG124-BI4-VG10.png} we display lateral averages of the total energy as functions of time.
			We find asymmetries in both quantities: initially, the thermal wave is biased to the right.
			Then, between $\tau (\times10^{-3}) =$ 0 and 5 there is an expansion regime where there is thermalization.
			For $\tau (\times10^{-3}) >5$ the daemon's action enters the game and the waves are segregated again.
			These curves are compared with those of fig.~\ref{fig:PsiVPsiE-SG124-BI4-VG10.png},
			where indeed the average potential energy becomes negative for $\tau (\times10^{-3}) >5$,
			indicating that the particles exert work on the daemon (see a global minimum of $\prom V$ at $\tau (\times10^{-3})\simeq 9$.
			In this setting, we conclude that our device operates well until the wave expansion allows an important interaction with $V$ at $\tau (\times10^{-3}) =5$ and after.
			For very large times, a regime with noisy «collapse-and-revival» can be seen.
			
			In fig.~\ref{fig:wBoltzPlotE-SG124-BI4-VG10.png} we show a density plot for \eqref{eq:PsiBoltz}.					
			The Talbot effect induces a recurrence time in the quasi-temporal coordinate that will force the system to repeat its behavior.
			In this case, $\tau_\text{Talbot}\simeq 10(\times10^3)$, and
			for $\tau <\tau_\text{Talbot}$ there is an asymmetry that shows the efficient work of the daemon.
			Subsequently the behavior is reversed between the compartments of the box.
			
			Another system of interest is the one is uniform, i.e. $\Psi_0^\text{I} = 1$,
			so the probability that the particle is in any position state is equally likely
			(this is denoted by $\beta=$IS0).
			For this case the entropy evolution is shown in fig.~\ref{fig:EntropyIS0-SG124-BI4-VG10.png}.
			Note that the entropy value oscillates, again reaching a minimum as the system evolves.
			A density plot of the wave function $\Psi_\text{I}(x,\tau)$ shown in fig.~\ref{fig:wBoltzPlotIS0-SG124-BI4-VG10.png},
			reveals that the wave is distributed asymmetrically due to terms that breaks parity explicitly as expected.

\section{Emulation of a Quantum Dynamical Daemon in Electromagnetic Billiards}\label{sec:EQD}
\begin{figure}[b]\centering		
		\includegraphics[width=0.45\textwidth]{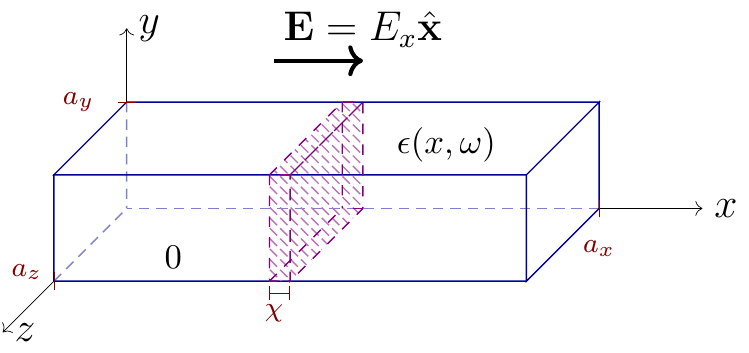}
		\caption{Rectangular cavity with sides $ (a_x, a_y, a_z) $ where a dielectric with permittivity
		$\epsilon (x,\omega)$ and thickness $\chi$ has been placed in the center.
		The electric field goes in the $\hat \vecx$ direction and the blue lines denote plates of a good conductor.}
		\label{fig:GrGO}
	\end{figure}

	The quantum daemon described in the previous section can be emulated in an electromagnetic cavity as in fig.~\ref{fig:GrGO}.
	Such a daemon is modeled as a dielectric in the middle of a conducting cavity with $\chi$ thickness,
	and reflects the radiation in the entire $yz$ plane.
	Therefore, the dielectric function that simulates the interaction must have a dependence only in $x$, i.e. $\epsilon (x,\omega)$.
	Using Maxwell's equations, the description of the electric field $\vecE$ inside a vacuum waveguide is given by
	\begin{align}\label{eq:JackyOchoCientoOcho}
		\nabla^2\vecE+\mu\omega^2\epsilon(\vecx,\omega)\vecE+\nabla\left(\frac{1}{\epsilon(\vecx,\omega)}\,\vecE\cdot\nabla\epsilon(\vecx,\omega)\right)=0.
	\end{align}
	Since the electric field only goes in the $x$-direction,
	and using the reduced dependence of the dielectric function,
	i.e. 1D-problem, the following simplification is obtained
	\begin{align}
		\partial_{xx}^2E_x+\mu_r\epsilon_r(x,\omega)\frac{\omega^2}{c^2}E_x
		+\partial_x\Bparen{E_x\partial_x\log\epsilon_r(x,\omega)}=0,
	\end{align}
	where the relative permittivity $ \epsilon_r (x, \omega) = \epsilon (x, \omega) / \epsilon_0 $
	and similarly for the relative magnetic permeability $\mu_r = \mu / \mu_0$,
	although this will remain a constant throughout our considerations.
	The above equation allows separable solutions $E_x (x,y,z) = \phi(y,z)\Psi(x)$,
	for which the perpendicular part $\phi(y,z)$ satisfies the Helmholtz equation
	with two different boundary conditions on the surface (Dirichlet for parallel field, Neumann for perpendicular field at $x =0, a_x$).
	The longitudinal part $\Psi(x)$ satisfies
	\begin{align}\label{eq:JackySchEx}\begin{autobreak}
		-\Psi''(x)+\xi^2\Psi(x)
		-\mu_r\epsilon_r(x,\omega)\frac{\omega^2}{c^2}\Psi(x)
		-\partial_x\Bparen{\Psi(x)\partial_x\log\epsilon_r(x,\omega)} =0.
	\end{autobreak}\end{align}	
	This equation can be viewed in two different ways:
	Given the $\epsilon_r$ function, $\Psi(x)$ can be determined employing the aforementioned boundary conditions,
	so the field $\vecE_x$ is determined.
	On the other hand, this equation can be compared with our Schrödinger theory for a daemon placed at the central part of the box,
	which forces to consider certain special relations for epsilon in the form of a new differential equation.
	We shall see this in the next section.
	
	\subsection{From Energies to Frequencies}	
	As we have seen, the left-right symmetry breaking produced by Maxwell's daemon in quantum mechanics
	can be mimicked by linear terms in the field's derivative,
	included in the Helmholtz wave operator (normal incidence of the field on the dielectric).
	However, the original mechanical setting required knowledge of the wave number $k$,
	whereas non-local dielectric functions depend on the frequency $\omega$ of the generated field
	(e.g. by the an antenna inside the cavity).
	It is possible to produce a similar daemonic formulation in electromagnetic settings,
	but this time with a frequency selector incorporated in the dielectric function,
	and as long as the asymmetry is determined by the behavior of $\epsilon(x,\omega)$ under parity exchange $x\rightarrow -x$.
	We have:
	\begin{equation}
		V(x,p) \mapsto V(x,\sgn p, E),
	\end{equation}
	and since $\omega = E/\hbar$, $k = p/\hbar$, we have a function $V(x,\sgn(k),\omega)$.
	
	The aim here is to obtain an explicit form of a dielectric function $\epsilon(x,\omega)$
	such that the form of the Helmholtz equation matches that of the Schrödinger equation with irreversible terms
	(interactions that break P and T symmetry).
	Substituting $ V_ \text {act} (p) $ for the Hermitian potential in \eqref{eq:PotHer},
	it is obtained from \eqref{eq:DdMQMEq} that
	\begin{align}\label{eq:SchVact}\begin{autobreak}
		\frac{2m}{\hbar^2}\Big(\hat V(\hatx,\hatp) -E\Big)
			\mapsto
			\frac{1}{2}\partial_x\frac{c f_-(\omega)}{i\omega}V_\text{B}(x)
			+\frac{1}{2}V_\text{B}(x)\frac{c f_-(\omega)}{i\omega}\partial_x
			+\frac{1}{2}\AConm{f_+(\omega)}{V_\text{B}(x)} -\frac{2m}{\hbar^2}E.
	\end{autobreak}\end{align}
	where the correspondence principle was used
	$\hatp\rightarrow -i\hbar\partial_x$ and $|p|=\hbar\omega/c$.	
	On the other hand, \eqref{eq:JackySchEx} can be transformed into
	\begin{align}\label{eq:JackySchPsi}\begin{autobreak}
		-\Psi''(x)
		+\xi^2\Psi(x)-\mu_r\epsilon_r(x,\omega)\frac{\omega^2}{c^2}\Psi(x)
		-\frac{1}{2}\Bparen{\partial^2_{xx}\log\epsilon_r(x,\omega)}\Psi(x)
		-\frac{1}{2}\left(\partial_x\Bparen{\partial_x\log\epsilon_r(x,\omega)}\right.
		+\left.\Bparen{\partial_x\log\epsilon_r(x,\omega)}\partial_x\right)\Psi(x)=0,
	\end{autobreak}\end{align}	
	and comparing with \eqref{eq:SchVact} it follows that
	\begin{subequations}
	\begin{align}
		V_\text{B}(\hatx) = -\frac{i\omega}{c f_-(\omega)}\partial_x\log\epsilon_r(x,\omega),
	\end{align}
	and
	\begin{align}\begin{autobreak}
		\frac{1}{2}\AConm{f_+(\omega)}{V_\text{B}(x)} -\frac{2m}{\hbar^2}E
			= \xi^2 -\mu_r\epsilon_r(x,\omega)\frac{\omega^2}{c^2}
				-\frac{1}{2}\partial^2_{xx}\Bparen{\log\epsilon_r(x,\omega)}.
	\end{autobreak}\end{align}
	\end{subequations}
	Likewise, separating the contribution of the vacuum and the medium according to
	$\epsilon_r(x,\omega) = 1 + \eta_r(x,\omega)$, it follows that
	\begin{subequations}	
	\begin{align}
		-\frac{2m}{\hbar^2}E = \xi^2-\mu_r\frac{\omega^2}{c^2}
	\end{align}
	and
	\begin{align}\label{eq:EDNLVact0}\begin{split}
		\mu_r\frac{\omega^2}{c^2}\Big(\epsilon_r(x,\omega)-1\Big)
			&+\frac{1}{2}\partial^2_{xx} \Bparen{\log\epsilon_r(x,\omega)} \\
		&= \frac{i\omega}{c} \cdot \frac{f_+(\omega)}{f_-(\omega)}
			\cdot \partial_x\Big(\log\epsilon_r(x,\omega)\Big).
	\end{split}\end{align}
	\end{subequations}
	Acceptable solutions to this differential equation for various initial conditions
	will lead to different types of embodiment that respect Kramers--Kronig causality conditions, as explained in \cite{Jackson1999}.
	Part of the methodology consists in solving \eqref{eq:JackySchEx} using vector mappings around fixed points.
	There are two regimes for the solution of the differential equation for electrical permittivity.
	In one case there are frequencies below the ultraviolet cut-off (i.e. the daemon operates),
	so we should compare \eqref{eq:JackySchEx} with eq. from Schrödinger with potential;
	while for the second case there are frequencies above the ultraviolet cut (i.e. the daemon stops operating),
	so that potential will be zero in comparison.
	
	It should be noted that all the analysis performed in this section can be greatly simplified by a harmonic approximation,
	leading to analytical results for this function. In order to do so, let
	\begin{equation}
		\log\epsilon_r(x,\omega) = v(x)\exp\left(\frac{i\omega}{c}\cdot\frac{f_+(\omega)}{f_-(\omega)} x\right),
	\end{equation}
	and \eqref{eq:EDNLVact0} now reads
	\begin{align}\begin{split}
		\partial^2_{xx} v(x) &- \left(\frac{i\omega}{c}\cdot\frac{f_+(\omega)}{f_-(\omega)} \right)^2 v(x) = \\
		&- 2\mu_r\left(\exp\Bparen{v(x)e^{\frac{i\omega}{c}\cdot\frac{f_+(\omega)}{f_-(\omega)} x}} -1\right)\frac{\omega^2}{c^2}
		e^{-\frac{i\omega}{c}\cdot\frac{f_+(\omega)}{f_-(\omega)} x}.
	\end{split}\end{align}
	and by expanding the exponential term to first order, we get
	\begin{align}\begin{split}
		v(x) &\simeq v_0\cos\left(x\sqrt{1+2\mu_r}\frac{\omega}{c}\right) \\
			&+ \frac{v'_0}{\sqrt{1+2\mu_r}}\frac{1}{\omega/c}\sin\left(x\sqrt{1+2\mu_r}\frac{\omega}{c}\right),
	\end{split}\end{align}
	where $v_0=v(0)$ is the initial condition.
	
	This equation represents good solutions to \eqref{eq:EDNLVact0} in a regime where $\epsilon(x,\omega)$ has small deviations from unity.
	
	\begin{figure}[htpb]
		\subfloat[\label{fig:ReEDNLYB}]{%
			\includegraphics[width=0.5\textwidth]{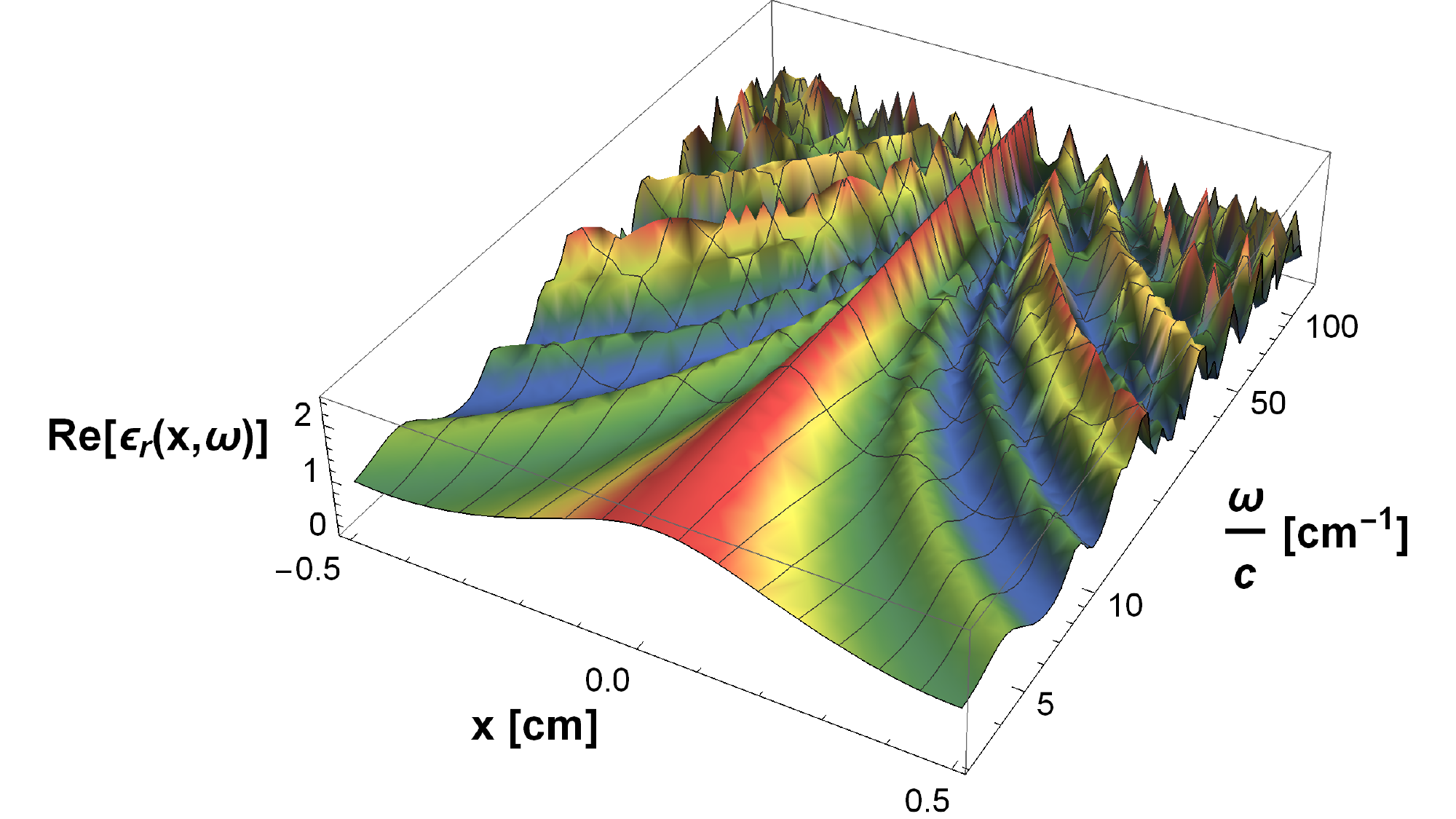}%
		}\hfill
		\subfloat[\label{fig:ImEDNLYB}]{%
			\includegraphics[width=0.5\textwidth]{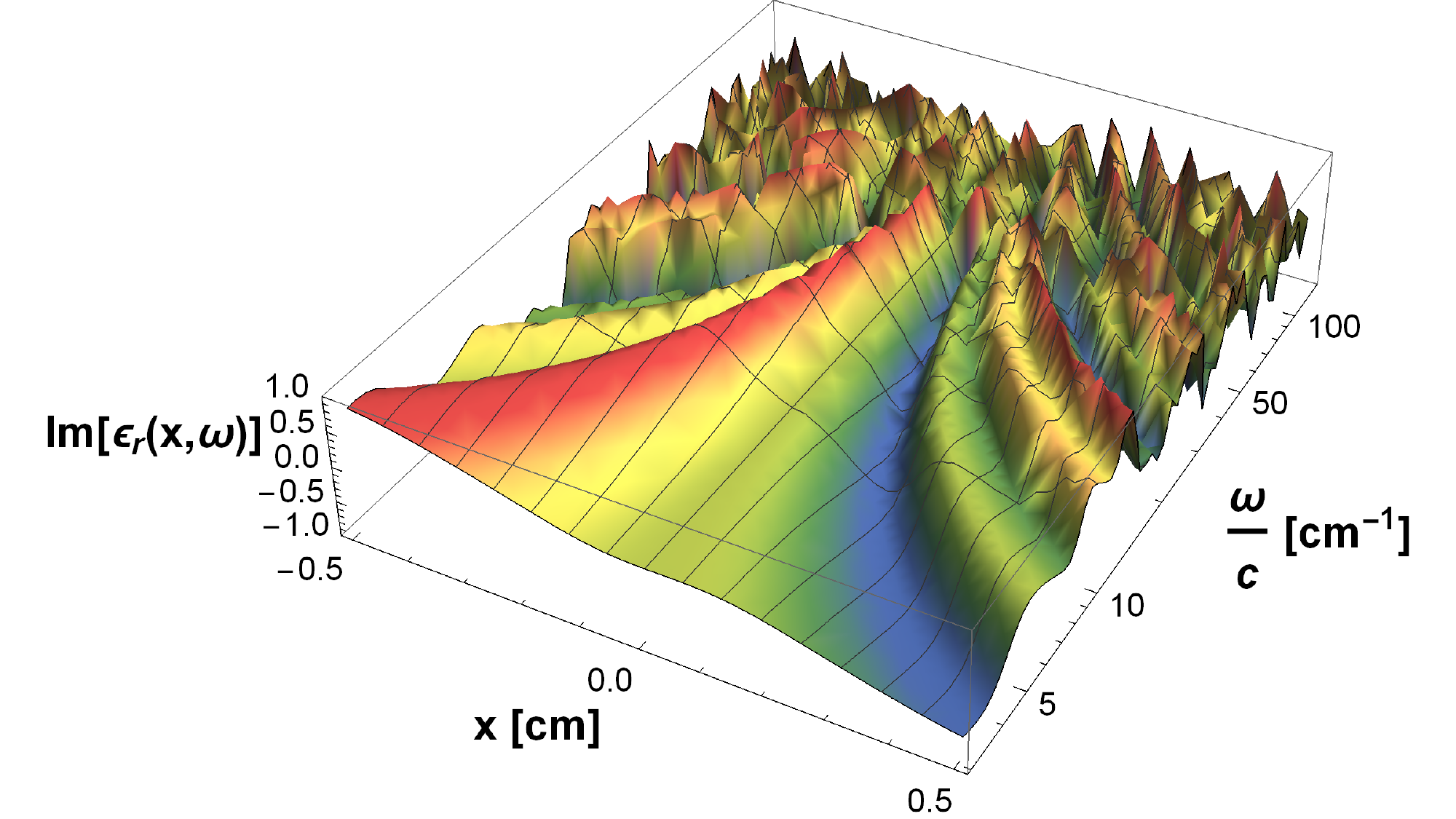}%
		}
       	\caption{a) Real and b) Imaginary parts of the solution to eq.\eqref{eq:EDNLVact0}
		with $\epsilon_r(0) =+2$ and $\epsilon'_r(0) =+2$ at $\pi\leq\omega/c\leq40\pi$.}  
		\label{fig:ReImEDNLYB}      
	\end{figure}%
	
	\subsection{Daemonic Dielectric Function as an Inverse Problem}
	Differential eq.\eqref{eq:EDNLVact0} is solved numerically with suitable initial conditions (or boundary values)
	\begin{align}
		\epsilon_r(0) =+2 && \text{and} && \epsilon'_r(0) =+1.
	\end{align}
	The value of $ \omega/c $ is varied from $\pi$ to $40\pi$ cm$^{-1}$ as a parameter,
	with $\omega_R/c=16\pi$ cm$^{-1}$.	
	In fig.~\ref{fig:ReImEDNLYB} we show the real and imaginary parts of the dielectric function
	needed to emulate Maxwell's daemon.
	It is found that the dielectric function is even in its real part and odd in its imaginary part.
	This behavior will be useful when designing the emulation of the electromagnetic cavity with an effective dielectric.

\section{Realization of an Electromagnetic Dynamical Daemon with Dielectric Slabs}\label{sec:EEDD}
\begin{figure}[b]\centering		
		\includegraphics[width=0.5\textwidth]{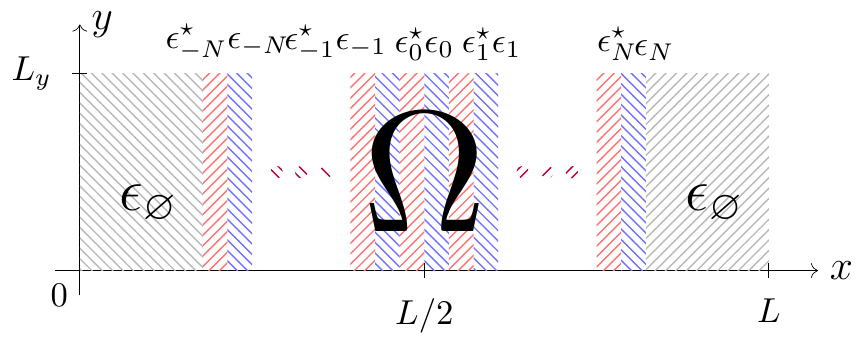}
		\caption{System formed by an ensemble $\Omega$ of $2N+1$ dielectric bilayers centered in a vacuum metal cavity ($\epsilon_\varnothing$) of $L$ length,
		with Neumann-type conditions on the walls ($x=0$ \& $x=L$)
			The tandem configuration of the slabs ensures that the real (imaginary) part of the dielectric function must be even (odd).}
		\label{fig:GSD}
	\end{figure}

	Here we proceed to emulate dynamically the propagation of a wave inside a cavity with judiciously designed dielectric functions.
	In order to do so, the wave function of the system given by \eqref{eq:JackySchEx} is decomposed into stationary modes.
	In this way it is obtained a non-linear (transcendental) eigenvalue problem of the form
	\begin{equation}\label{eq:Dxwx}
		\mathcal{D}(\omega)\vecx = \frac{\omega^2}{c^2}\vecx.
	\end{equation}
	Using an orthonormal basis for a metal cavity over the interval $x\in[0,L]$ with Neumann-type conditions on the walls ($x=0$ \& $x=L$), i.e.
	\begin{equation}
		\phi_m(x)=\sqrt{\frac{2}{L}}\,\cos\big(\kappa_m x\big)
	\end{equation}
	with
	\begin{equation}
		\kappa_m = \frac{m\pi}{L}, \qquad\text{\&}\qquad m = 0, 1, 2, \ldots,
	\end{equation}
	the matrix elements $\bra{m}\mathcal{D}(\omega)\ket{n}$ are calculated as
	\begin{align}\label{eq:Dxwxint}\begin{autobreak}
		\bra{m}\mathcal{D}(\omega)\ket{n} =
		- \int dx \, \phi_m(x) \Bigg\lbrace \partial^2_x - \xi^2
		+ \mu_r\epsilon_r(x,\omega)\frac{\omega^2}{c^2}
		+ \partial^2_x\log\epsilon_r(x,\omega) + \partial_x\log\epsilon_r(x,\omega)\partial_x \Bigg\rbrace \phi_n(x).
	\end{autobreak}\end{align}
	(See appendix \ref{ap:c} for full expressions.)
	At the center of this cavity --initially in a vacuum ($\epsilon_\varnothing$)--
	an ensemble $\Omega$ of $2N+1$ contiguous bilayer dielectric curtains is added (see fig.~\ref{fig:GSD}).
	Such a tandem configuration of the slabs ensures that the real (imaginary) part of the dielectric function
	must be even (odd) according to the solutions found previously for the dielectric permittivity of the daemon.
	Therefore, the right (left) part of the $k$-th curtain $\epsilon_k$ ($\epsilon_k^\star$) is modeled according to Lorentz--Drude as
	\begin{equation}
		\epsilon_k(\omega,\Gamma_k) =
		1 + \frac{\omega^2_{p_k}}{\omega^2_{0_k} - \omega^2 - i\Gamma_k\omega} =
		\epsilon_k^\star(\omega,-\Gamma_k),
	\end{equation}
	whose spring, plasma and damping constants are $\omega_0$, $\omega_p$ and $\Gamma$, respectively.
	(The $\star$ notation is employed to indicate items corresponding to the left side of the bilayer,
	not to be confused with complex conjugate $\epsilon_k^*$). 
	Therefore, the dielectric function of the system ($0\leq x\leq L$) is given by
	\begin{subequations}
	\begin{align}\begin{split}
		\epsilon_r\bparen{ x,\omega } =
		\Theta\left( x -a_{N+\frac{1}{2}} \right) &+ \Theta\left( a_{-N-\frac{1}{2}} -x \right) \\
		+ \sum_{k=-N}^N &\Big\{\epsilon_k(\omega,\Gamma_k)\cdot\sqcap_k(x,a_k) \\
		&+ \epsilon_k^\star(\omega,\Gamma_k^\star)\cdot\sqcap_k^\star(x,a_k)\Big\}
	\end{split}\end{align}
	where $\sqcap_k$ is a rectangular window of thickness $b/2$ in the position $a_k$
	\begin{equation}
		\sqcap_k(x,a_k) = \Theta\bparen{x -a_k} - \Theta\left(x -a_k -\frac{b}{2}\right) = \sqcap_k^\star(-x,-a_k),
	\end{equation}
	with
	\begin{equation}
		a_k = \frac{L}{2} + kb.
	\end{equation}
	\end{subequations}
	It should be mentioned that the derivative of the term $\epsilon_r$ is interpreted --according to distribution theory-- as
	\begin{equation}
		\partial_x\log\epsilon_r = \begin{cases}
		\displaystyle \sum_{k=-N}^N \Big\lbrace \mathrm{log}\,\epsilon_k(\omega,\Gamma_k)\cdot\partial_x\sqcap_k(x,a_k)  \\
			\hfill + \;\star\; \Big\rbrace \hfill \text{ if } x\in\Omega, \\
		~\\ \hfill \displaystyle 0 \hfill \text{ if } x\notin\Omega, \end{cases}
	\end{equation}
	using the principal branch of the logarithm
	and $\star$ represents the corresponding terms for $\epsilon_k^\star$.
	
	In order to reconstruct the wave function in time,
	we consider its decomposition in eigenmodes of the empty cavity, denoted by $n$,
	for each of the eigenfrequencies $\omega_m$ of the cavity with dielectrics as
	\begin{equation}\label{eq:PsiModos}
		\braket{x}{\Psi,t} = \sum_{m,n}
		\braket{x}{n}\braket{n}{\omega_m}\braket{\omega_m}{\Psi_0}\exp\bparen{-i\omega_m t}
	\end{equation}
	where $\braket{x}{n} = \phi_n(x)$ is the Neumann basis,
	the eigenvectors $\ket{\omega_m}$ have components $\braket{n}{\omega_m} = v_n(\omega_m)$
	and the constants $\braket{\omega_m}{\Psi_0} = c_m$
	constitute the overlap of the basis $\phi_n(x)$ with the initial wave function $\Psi(x,0)$ in the form
	\begin{equation}
		\sum_m v_n(\omega_m)c_m = \int_0^L dx\, \phi_n(x) \Psi(x,0).
	\end{equation}
	\begin{subequations}
	The above decomposition into eigenmodes is expressed in matrix form as follows:
	The basis is rewritten as a vector $\Phi(x)$ of entries $\phi_n(x)$, i.e.
	\begin{equation}
		\phi_n(x) \;\rightarrow\; \Phi_{n\times 1}(x) = \Bparen{ \phi_0(x), \phi_1(x), \cdots, \phi_{n-1}(x) }^T.
	\end{equation}
	The eigenvectors $v_n(\omega_m)$ now constitute the rows of the rectangular matrix $\mathbb{V}$, i.e.
	\begin{equation}\resizebox{0.9\hsize}{!}{$
		v_n(\omega_m) \;\rightarrow\; \mathbb{V}_{m\times n} = \begin{pmatrix}
		v_0(\omega_1) & v_1(\omega_1) & \cdots & v_{n-1}(\omega_1) \\
		v_0(\omega_2) & v_1(\omega_2) & \cdots & v_{n-1}(\omega_2) \\
		\vdots & \vdots & \ddots & \vdots \\
		v_0(\omega_m) & v_1(\omega_m) & \cdots & v_{n-1}(\omega_m) \end{pmatrix}
	$}\end{equation}
	The exponentials $\exp\bparen{i\omega_m t}$ conform the diagonal matrix $\mathbb{E}\mathbb(t)$, i.e.
	\begin{align}\begin{split}
		\exp\bparen{i\omega_m t} \;&\rightarrow\; \mathbb{E}_{m\times m}(t) \\
			&= \mathrm{diag}\Bparen{ \exp\bparen{i\omega_1 t}, \cdots, \exp\bparen{i\omega_m t} }.
	\end{split}\end{align}
	The constants $c_m$ form a covector $\mathbf{c}$,
	made of a product of the initial vector $\mathbf{b}$ (whose components represent the overlap of the initial wave function with the basis),
	with the inverse of the eigenvector matrix, i.e.
	\begin{equation}
		c_m \quad\rightarrow\quad \mathbf{c}_{1\times m} = \mathbf{b}_{1\times n}\mathbb{V}^{+}_{n\times m},
	\end{equation}
	where we made use of the generalized inverse \citep{RTGregory2011,AdiBenIsrael2003}
	\begin{equation}
		\mathbb{V}^{+} = \Bparen{\mathbb{V}^\dagger \mathbb{V}}^{-1} \mathbb{V}^\dagger.
	\end{equation}
	\end{subequations}
	Finally, the wave function is obtained as
	\begin{equation}
		\Psi(x,t) = \mathbf{c}_{1\times m} \mathbb{E}_{m\times m}(t) \mathbb{V}_{m\times n} \Phi_{n\times 1}(x).
	\end{equation}
	(See appendix \ref{ap:c} for an explanatory note on the method for obtaining $v_n(\omega_m)$ and $\omega_m$.)
	
	\begin{figure}[b]\centering
	\includegraphics[width=0.5\textwidth]{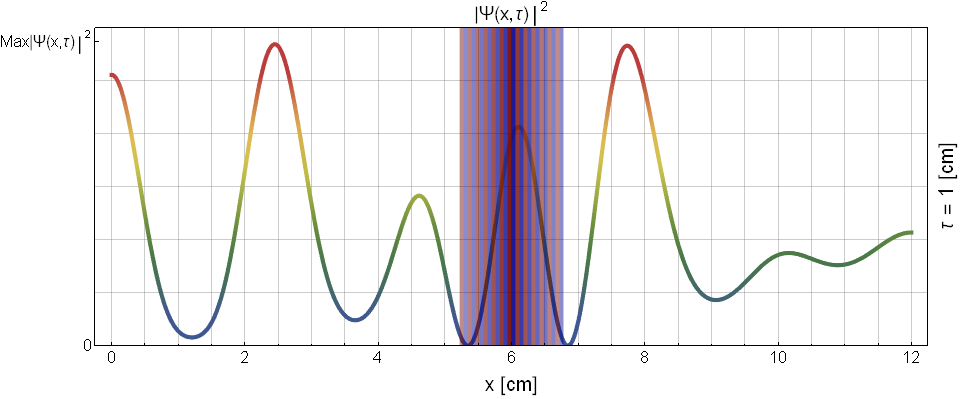}
	\caption{Wave function profile at $\tau= 1$ cm with 12 modes.
	The system is composed of 13 dielectric bilayers, each with a thickness of $b=1.2$ mm, centered in a cavity of $L=12$ cm.}
	\label{fig:FOPN7-C13-M12-0}
	\end{figure}%
	
	\begin{figure}[htpb]
		\subfloat[\label{fig:FDPN7-C13-M12-1}]{%
			\includegraphics[width=0.5\textwidth]{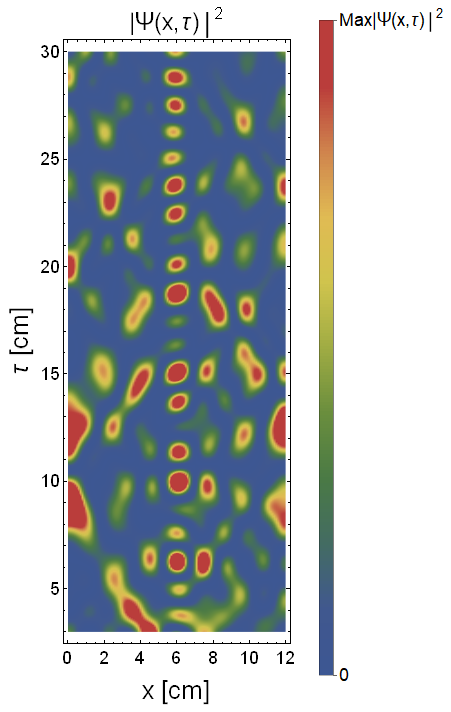}%
		}\hfill
		\subfloat[\label{fig:FDPN7-C13-M12-0}]{%
			\includegraphics[width=0.5\textwidth]{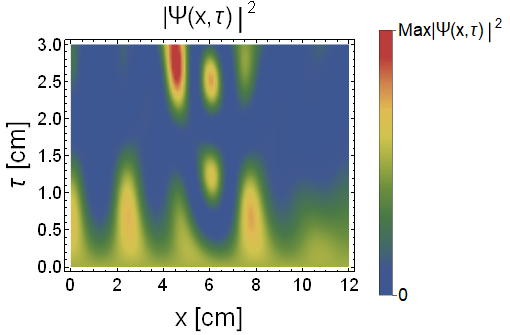}%
		}
       	\caption{Wave function density graph with 12 modes at a) $0\leq\tau\,\text{[cm]}\leq 30$, b) $0\leq\tau\,\text{[cm]}\leq 3$.
		The system is composed of 12 dielectric bilayers, each with a thickness of $b=1.2$ mm, centered in a cavity of $L=12$ cm.}  
		\label{fig:FDPN7-C13-M12}      
	\end{figure}%
	
	\subsection{Results of the Emulation}
	For simulation purposes,
	we introduce 13 dielectric bilayers such that
	$\epsilon_k(\omega)=\epsilon_{-k}(\omega)$ \& $\epsilon^\star_k(\omega,\Gamma) = \epsilon_k(\omega,-\Gamma)$.
	In table \ref{tab:tbc} we find numerical values of parameters employed in the construction of the dielectric bilayers
	(these values have been rescaled with respect to $c$, so the frequency units are the inverse of the distance,
	consequently, the time $\tau$ in the wave function becomes in units of distance).
	\begin{table}[h]\centering
		\begin{tabular}{c|ccc}
		\hline
		$k$-th bilayer & $\displaystyle\widetilde{\omega_p}$ [cm$^{-1}$] & $\displaystyle\widetilde{\omega_0}$ [cm$^{-1}$] & $\displaystyle\widetilde{\Gamma}$ [cm$^{-1}$] \\
		\hline
		$0$ 	& \multirow{3}[0]{*}{$\vdots$} & 1.20 & 0.050 \\
		$\pm 1$ & \multirow{5}[0]{*}{3} & 1.08 & 0.045 \\
		$\pm 2$ &  & 0.96 & 0.040 \\
		$\pm 3$ & \multirow{4}[0]{*}{$\vdots$} & 0.84 & 0.035 \\
		$\pm 4$ &  & 0.72 & 0.030 \\
		$\pm 5$ &  & 0.60 & 0.025 \\
		$\pm 6$ &  & 0.48 & 0.020 \\
		\hline
		\end{tabular}
	\caption{Dielectric curtain constants with the conditions
	$\epsilon_k(\omega)=\epsilon_{-k}(\omega)$ \& $\epsilon^\star_k(\omega,\Gamma) = \epsilon_k(\omega,-\Gamma)$.
	The tilde represents that the frequencies have been rescaled with respect to $c$ so that the units are the inverse of the distance (cm$^{-1}$),
	The thickness of each bilayer is $b=1.2$ mm in a cavity of $L=12$ cm.}
	\label{tab:tbc}
	\end{table}
	Also, each bilayer has a thickness of $b=1.2$ mm and the full system is contained in a cavity of length $L=12$ cm.
	For the case of uniformly distributed waves as initial condition we have the vector $\mathbf{b} = (1,0,\ldots,0)$.
	Using 12 modes (who are in the zone where the daemon sees and operates over them),
	an asymmetry in the wave function is shown in $\tau= 1$ cm (or $t=33$ ps) in fig.~\ref{fig:FOPN7-C13-M12-0}.
	In fig.~\ref{fig:FDPN7-C13-M12-0} there is a density plot at short times.
	Although the asymmetries constitute a transient effect,
	a predominant behavior is observed on the right compartment of the box as time evolves further,
	including several asymmetry windows like the one shown in fig.~\ref{fig:FDPN7-C13-M12-1}.

\section{Conclusions}
	A dynamical model for a system that splits an ensemble of waves representing independent particles has been proposed and successfully studied.
	The so-called dynamical Maxwell's daemon --thus obtained--
	works with a reference momentum that helps to decide how two subsystems with different temperatures are distributed in two compartments of a cavity.
	Our description has been possible via a novel Hamiltonian operator given by \eqref{eq:HamV} and the irreversible potential in \eqref{eq:VactSgn}.
	The outcome is reminiscent of the classical daemon's action shown in fig.~\ref{fig:EFEnsamble},
	as we have confirmed by analyzing wave dynamics in fig.~\ref{fig:wBoltzPlotE-SG124-BI4-VG10.png}.
	As an outstanding result, the undulatory version of Maxwell's daemon contains --in its evolution--
	the interference structure of Talbot (quantum) carpets in time.
	For long times, a collapse and revival structure can be distinguish.
	This structure displays the expected asymmetries for limited periods of time associated with Talbot lengths.
	There is no true thermalization (as opposed to the classical process) because of such revivals. 

	We would like to stress that most of the results have been obtained analytically;
	in particular we have reported in this study a new asymmetrical Green's function pertaining to irreversible systems,
	not found in standard references \cite{GroscheSteiner1998}.
	The meromorphic structure of such analytical answer has been docile enough to allow proper identification of energy ranges where the device is effectual.

	Likewise, the dynamics of the daemon were emulated by employing Maxwell's equations with a spatial and frequency dependence in the dielectric function,
	allowing to design a metamaterial by layer juxtaposition with the sought properties.

	A number of quantities and their time behavior support our conclusions in connection with irreversibility and the apparent entropy decrease.
	Indeed, with Shannon's definition for a basis-dependent disorder function (energy states)
	we observe regimes where ordered configurations are established as time elapses.
	Also, densities and average energies at each compartment were studied.
	(Figs.~\ref{fig:PsiHPsiE-SG124-BI4-VG10.png} and \ref{fig:FDPN7-C13-M12} are unmistakable in this sense.)

	In general, we could say that the second law of thermodynamics might be restored if the average work done by the potential is taken into account,
	instead of the entropic evolution of the waves alone.
	This old resolution of the paradox in classical physics also finds its way in quantum dynamics without collapse postulates.
	In contrast, other explanations offered in the past rest heavily on information theory,
	as long as measurements are also part of the evolution process.

	From here, a very strong result follows: our construction of Maxwell's daemon can be applied to any system of waves,
	with the aim of cooling down radiation (e.g. electromagnetic waves, acoustic and elastic waves) by separating «hot» and «cold» components in space.
	In this direction, we have studied an electromagnetic system with mode-decomposition techniques,
	with the aim of predicting its general behavior in a realistic construction.

%
%
%
%
%
\appendix

\section{\label{ap:a} The Green's Function}
	Prior to solving \eqref{eq:GreenPotencialDelta}, we start with
	\begin{equation}
		\Bparen{H-E +V_0\delta(\hatx)} \hat G_\delta = \mathbb{I}.
	\end{equation}
	Multiplying from left with $\hat G_0$ and computing it in the position basis $x$
	\begin{equation}
		G_\delta(x,x',E) + V_0\bra{x} \hat G_0\delta(\hatx) \hat G_\delta\ket{x'} = G_0(x,x',E),
	\end{equation}
	where it has been used that
	\begin{equation}
		\Bparen{H-E} \hat G_0 = \mathbb{I},
		\quad\text{and}\quad \bra{x} \hat G_0 \ket{x'} = G_0(x,x',E).
	\end{equation}	
	Inserting a continuous complete set, obtains
	\begin{align}\label{eq:GreenDeltaInt}\begin{split}
		G_\delta(x,x',E) &+ V_0\int dx'' G_0(x,x'',E) \delta(x'') G_\delta(x'',x',E) \\
			&= G_0(x,x',E).
	\end{split}\end{align}
	The above expression is evaluated at $x=0$, yielding a functional equation
	\begin{equation}
		G_\delta(0,x',E) = \frac{G_0(0,x',E)}{1+V_0 G_0(0,0,E)},
	\end{equation}
	which finally, when introduced in \eqref{eq:GreenDeltaInt}, obtains
	\begin{equation}\label{eq:GreenDelta}
		G_\delta(x,x',E) = G_0(x,x',E) - \frac{V_0G_0(x,0,E)G_0(0,x',E)}{1+V_0 G_0(0,0,E)}.
	\end{equation}
	Now a momentum-dependent potential is added,
	so the equation to be satisfied is
	\begin{equation}
		\Bparen{H-E +V(\hatp)\delta(\hatx)+\delta(\hatx)V(\hatp)} \hat G_p = \mathbb{I}.
	\end{equation}
	Operating as in \eqref{eq:GreenDeltaInt}, we get for this case
	\begin{align}\label{eq:GreenPInt}\begin{split}
		G_p(x,x',E)& = G_0(x,x',E) \\
			&- \int dy\,G_0(x,y,E)\delta(y)V(\hatp)G_p(y,x',E) \\
			&- \int dy\,G_0(x,y,E)V(\hatp)\delta(y)G_p(y,x',E).				
	\end{split}\end{align}
	where the operator $\hat p$ indicates the momentum-dependence and is understood as $-i\hbar\partial_y$.
	Prior to the evaluation of the above expression,
	we insert another complete set in each integral in the form
	\begin{subequations}\label{eq:GreenPInt2}
		\begin{align}\begin{split}
			& \bra{y}\delta(\hat x)V(\hatp)\hat G_p \ket{x'}
				= \int dy'\,\bra{y}\delta(\hat x)V(\hatp)\ketbra{y'}{y'} \hat G_p \ket{x'} \\
				&= \int dy'\,\frac{\delta(y)}{2\pi}\int dp\,e^{ip(y-y')}V(p)G_p(y',x',E) \\
				&= \delta(y)\int dy'\,\widetilde V(y')G_p(y',x',E),
		\end{split}\end{align}
		where a complete set of plane waves was introduced in the second line, and
		\begin{equation}
			\widetilde V(y') = \frac{1}{2\pi}\int dp\, e^{-ipy'}V(p),
		\end{equation}
		is the Fourier transform of the potential;
		whereas, for the second integral
		\begin{align}\begin{split}
			& \bra{y}V(\hatp)\delta(\hat x)\hat G_p \ket{x'}
				= \int dy'\,\bra{y}V(\hatp)\delta(\hat x)\ketbra{y'}{y'} \hat G_p \ket{x'} \\
				&= \int dy'\,\frac{\delta(y')}{2\pi}\int dp\,e^{ip(y-y')}V(p)G_p(y',x',E) \\
				&= \widetilde V(-y)\int dy'\,\delta(y')G_p(y',x',E) \\
				&= \widetilde V(-y) G_p(0,x',E).
		\end{split}\end{align}
	\end{subequations}
	Substitution of \eqref{eq:GreenPInt2} in \eqref{eq:GreenPInt}, leads to
	\begin{align}\begin{split}
		&G_p(x,x',E) = G_0(x,x',E) \\
			&- G_0(x,0,E)\int dy'\,\widetilde V(y')G_p(y',x',E) \\
			&- \int dy\,G_0(x,y,E)\widetilde V(-y) G_p(0,x',E).
	\end{split}\end{align}
	In order to get $G_p(x,x',E)$, we first multiply the last expression by $\widetilde V(x)$ and integrate over $x$,
	\begin{align}\begin{split}
		&\int dx\,\widetilde V(x)G_p(x,x',E) = \quad \int dx\,\widetilde V(x)G_0(x,x',E) \\
			&- \int dx\,\widetilde V(x)G_0(x,0,E)\int dy'\,\widetilde V(y')G_p(y',x',E) \\
			&- \int dx\,\widetilde V(x) \int dy\,G_0(x,y,E)\widetilde V(-y) G_p(0,x',E),
	\end{split}\end{align}
	and recognizing that the integral on the left-hand side is the same as the one in the 2nd term of the right-hand side
	(with another integration variable), we solve for it in order to obtain
	\begin{equation}
		\int dy'\,\widetilde V(y')G_p(y',x',E) = \frac{P_1(x',E)-Q_1(E)G_p(0,x',E)}{1+Q_2(E)}.
	\end{equation}
	where $P_1(x',E)$, $P_2(x,E)$, $Q_1(E)$, $Q_2(E)$ were defined earlier in section III.A.
	Substituting the above equation into $G_p(x,x',E)$ given by \eqref{eq:GreenPInt}, obtains
	\begin{align}\begin{split}
		&G_p(x,x',E) = G_0(x,x',E) - P_2(x,E)G_p(0,x',E) \\
			&- G_0(x,0,E)\frac{P_1(x',E)-Q_1(E)G_p(0,x',E)}{1+Q_2(E)},
	\end{split}\end{align}
	The last equation is not yet a closed formula for $G_p$, for it depends on $G_p$ again.
	Evaluating at $x=0$ provides the functional equation
	\begin{align}\begin{split}
		&G_p(0,x',E) = G_0(0,x',E) - P_2(0,E)G_p(0,x',E) \\
			&- G_0(0,0,E)\frac{P_1(x',E)-Q_1(E)G_p(0,x',E)}{1+Q_2(E)},
	\end{split}\end{align}
	which can be solved for $G_p(0,x',E)$, leaving
	\begin{subequations}
		\begin{equation}
			G_p(0,x',E) = \frac{G_0(0,x',E)-G_0(0,0,E)R_1(x',E)}{1+P_2(0,E)-G_0(0,0,E)Q_3(E)},
		\end{equation}
		where
		\begin{align}
			R_1(x',E) = \frac{P_1(x',E)}{1+Q_2(E)}, && Q_3(E) = \frac{Q_1(E)}{1+Q_2(E)}.
		\end{align}
	\end{subequations}
	The latter equation finally is substituted into $G_p(x,x',E)$,
	to obtain its final expression in terms of $G_0(x,x',E)$,
	as in \eqref{eq:GreenPotencialDelta} in the text.

\section{\label{ap:b} Sine--integral Approximation and Meromorphic Structure}	
	In order to analyze the obtained Green's function,
	the integrals can be approximated, in particular,
	for a container the terms to be obtained are
	\paragraph{The Fourier transform $\widetilde V(y)$}
	\begin{align}\begin{split}
		\widetilde V(\pm y) &= \frac{1}{2\pi}\int dp\,e^{\mp i(p+i\epsilon)y}V(p) \\
			&= \frac{1}{2\pi}\left(\int_{-\infty}^{-P_R} + \int_{0}^{P_R} \right)dp\, e^{\mp i(p+i\epsilon)y} \\
			&\overset{\epsilon\rightarrow 0}{=} \frac{\pm 1}{2i\pi y}\Bparen{1-2\cos P_R y}.
	\end{split}\end{align}
	
	\paragraph{Integral $P_1^\text{C}(x',E)$}
	\begin{align}\begin{split}
		&P_1^\text{C}(x',E) = \int_{-L/2}^{L/2} dx\,\widetilde V(x)G_0^\text{C}(x,x',E) \\
			&= -\frac{2}{i\pi L}
			\sum_{n=1}\frac{\Si\bparen{\xi_+}-\Si\bparen{\xi_-}-\Si(n\pi)}{E_{2n}-E}\sin(\kappa_{2n}x'),
	\end{split}\end{align}
	where $\xi_\pm = \bparen{P_R\pm\kappa_{2n}}\frac{L}{2}$.
	
	\paragraph{Integral $Q_1^\text{C}(E)$}
	\begin{align}\begin{split}
		Q_1^\text{C}(E) &= \int_{-L/2}^{L/2} dx\,\widetilde V(x)\int_{-L/2}^{L/2} dy\,G_0^\text{C}(x,y,E)\widetilde V(-y)\\
			&=\frac{2}{\pi^2L}\sum_{n=1} \frac{\Bparen{\Si\bparen{\xi_+}-\Si\bparen{\xi_-}-\Si(n\pi) }^2}{E_{2n}-E}
	\end{split}\end{align}
	
	\paragraph{Integral $P_2^\text{C}(x,E)$}
	\begin{align}\begin{split}
		P_2^\text{C}(x,E) &= \int dy\,G_0^\text{C}(x,y,E)\widetilde V(-y) \\
			&= \frac{2}{i\pi L}
			\sum_{n=1}\frac{\Si\bparen{\xi_+}-\Si\bparen{\xi_-}-\Si(n\pi)}{E_{2n}-E}\sin(\kappa_{2n}x) \\
			&= - P_1^\text{C}(x,E).
	\end{split}\end{align}
	
	Given the behavior of the function $\Si$, the following approximation can be made:
	The argument is written in the form
	\begin{align}\begin{autobreak}
		\Si\bparen{n\pi+a} - \Si\bparen{n\pi-a}
		= \Si\bparen{\pi(n+\pen{a/\pi})+\pi\epsilon}
		- \Si\bparen{\pi(n-\pen{a/\pi})-\pi\epsilon}
	\end{autobreak}\end{align}
	where $\pen{a/\pi}$represents the integer part and $\epsilon$ the fractional part of $a/\pi$.
	Expanding around $\epsilon=0$,
	\begin{align}\begin{split}
		&\Si\bparen{n\pi+a} - \Si\bparen{n\pi-a} \simeq \\
		&\Si\bparen{\pi(n+\pen{a/\pi})} - \Si\bparen{\pi(n-\pen{a/\pi})} \\
		&\qquad +
		\epsilon\left(\frac{\sin\bparen{\pi(n+\pen{a/\pi})}}{\bparen{n+\pen{a/\pi}}} + \frac{\sin\bparen{\pi(n-\pen{a/\pi})}}{\bparen{n-\pen{a/\pi}}}\right) \\
		&\simeq \frac{\pi}{2}-\frac{\pi}{2}\Theta\bparen{n-\pen{a/\pi}}+\frac{\pi}{2}\Theta\bparen{\pen{a/\pi}-n}+\pi\epsilon\,\delta_{n,\pen{a/\pi}}.
	\end{split}\end{align}
	(Note: Given the original function, the step function is zero for the case $n=\pen{a/\pi}$.)		
	Thus, it follows that
	\begin{align}\begin{split}
		\sum_{n=1}^\infty &\frac{\Si\bparen{\xi_+}-\Si\bparen{\xi_-}}{E_{2n}-E}\\
			&\simeq \sum_{n=1}^{\pen{a/\pi}-1}\frac{\pi}{E_{2n}-E} + \frac{\pi}{2}\sum_{n=1}^\infty\frac{\delta_{n,\pen{a/\pi}}}{E_{2n}-E},
	\end{split}\end{align}
	and
	\begin{align}\begin{split}
		\sum_{n=1}^\infty\frac{\Si\bparen{n\pi}}{E_{2n}-E} \simeq \frac{\pi}{2}\sum_{n=1}^\infty\frac{1}{E_{2n}-E}.
	\end{split}\end{align}
	While
	\begin{align}\begin{split}
		\sum_{n=1}^\infty &\frac{\Bparen{\Si\bparen{\xi_+}-\Si\bparen{\xi_-}-\Si(n\pi) }^2}{E_{2n}-E}\\
			&\simeq \frac{\pi^2}{4}\sum_{n=1}^\infty\frac{1}{E_{2n}-E}
			- \frac{\pi^2}{4}\sum_{n=1}^\infty \frac{\delta_{n,\pen{a/\pi}}}{E_{2n}-E}.
	\end{split}\end{align}

\section{\label{ap:c} Numerical Method for Field Resolution with Non-linear Frequency Dependence}
	The integrals required to obtain the matrix elements in \eqref{eq:Dxwxint} are
	\begin{equation}
		I_1 = \int_0^L dx\;\phi_m(x)\partial^2_x\phi_n(x) = -\kappa_n^2\delta_{mn},
	\end{equation}
	\begin{equation}\label{eq:I2}
		I_2 = \int_0^L dx\,\phi_m(x)\xi^2\phi_n(x) = \delta_{mn}\Bparen{1 +\delta_{n,0}} \xi^2,
	\end{equation}
	with
	\begin{equation}
		\xi^2 = \left(\frac{n_y\pi}{L_y}\right)^2 + \left(\frac{n_z\pi}{L_z}\right)^2.
	\end{equation}
	\begin{widetext}
		\begin{align}\begin{split}
			I_3(x\in\Omega) & = \int_\Omega dx\; \phi_m(x)\epsilon_r(x,\omega)\phi_n(x) =	\\
			& \sum_{k=-N}^N \begin{cases} \displaystyle \frac{2}{(m^2-n^2)\pi}\left\lbrace
			\epsilon_k\cdot f_{m,n}^{(-)}\left(a_{k+\frac{1}{2}}\right) - \epsilon_k^\star\cdot f_{m,n}^{(-)}\left( a_{k-\frac{1}{2}}\right)
			- (\epsilon_k -\epsilon_k^\star)\cdot f_{m,n}^{(-)}(a_k)\right\rbrace & \text{ si } m\neq n \\ ~\\
			\displaystyle (\epsilon_k +\epsilon_k^\star)\frac{b}{2L} + \frac{1}{n\pi}\left\lbrace\epsilon_k\cdot g_{n,n}\left( a_{k+\frac{1}{2}}\right) - \epsilon_k^\star\cdot g_{n,n}\left( a_{k-\frac{1}{2}}\right) - (\epsilon_k -\epsilon_k^\star)\cdot g_{n,n}(a_k)\right\rbrace
			& \text{ si } m = n \neq 0 \\ \\
			\displaystyle (\epsilon_k +\epsilon_k^\star)\frac{b}{L} & \text{ si } m = n = 0 \end{cases}
		\end{split}\end{align}
		where
		\begin{align}
			f_{m,n}^{(\pm)}(u) = m\sin(\kappa_m u)\cos(\kappa_n u)\pm n\sin(\kappa_n u)\cos(\kappa_m u),
			&& \& && g_{m,n}(u) = \cos(\kappa_m u)\sin(\kappa_n u).
		\end{align}
		\begin{equation}
			I_3(x\notin\Omega) = \int_{\Omega^C} dx\;\phi_m(x)\epsilon_r(x,\omega)\phi_n(x) = \begin{cases} \displaystyle
			\frac{2}{(m^2-n^2)\pi}\left\lbrace f_{m,n}^{(-)}\left( a_{-N-\frac{1}{2}}\right) - f_{m,n}^{(-)}\left(a_{N+\frac{1}{2}}\right) \right\rbrace
			& \text{ si } m\neq n \\ \\
			\displaystyle 1 - (2N+1)\frac{b}{L} + \frac{1}{n\pi}\left\lbrace g_{n,n}\left( a_{-N-\frac{1}{2}}\right) - g_{n,n}\left(a_{N+\frac{1}{2}}\right) \right\rbrace
			& \text{ si } m = n \neq 0 \\ \\
			\displaystyle 2 - 2(2N+1)\frac{b}{L} & \text{ si } m = n = 0 \end{cases}
		\end{equation}
		\begin{align}\begin{split}
			I_4 = \int_0^L dx\; \phi_m(x) & \partial^2_x\log\epsilon_r(x,\omega)\phi_n(x) = \\
			- & \frac{2\pi}{L^2}\sum_{k=-N}^N \left\lbrace\log\epsilon_k\cdot f_{m,n}^{(+)}\left( a_{k+\frac{1}{2}}\right)
			- \log\epsilon_k^\star\cdot f_{m,n}^{(+)}\left( a_{k-\frac{1}{2}}\right)
			- \bparen{\log\epsilon_k -\log\epsilon_k^\star }\cdot f_{m,n}^{(+)}(a_k)\right\rbrace.
		\end{split}\end{align}
		\begin{align}\begin{split}
			I_5 = \int_0^L dx\; \phi_m(x) & \partial_x\log\epsilon_r(x,\omega)\partial_x\phi_n(x) = \\
			+ & \frac{2n\pi}{L^2}\sum_{k=-N}^N \left\lbrace\log\epsilon_k\cdot g_{m,n}\left( a_{k+\frac{1}{2}}\right)
			- \log\epsilon_k^\star\cdot g_{m,n}\left( a_{k-\frac{1}{2}}\right)
			- \bparen{\log\epsilon_k -\log\epsilon_k^\star }\cdot g_{m,n}(a_k)\right\rbrace.
		\end{split}\end{align}
	\end{widetext}
	Since the elements of the $\mathcal{D}(\omega)$ matrix are made of a linear combination of algebraic
	($I_1$, $I_2$, $I_3$) and elementary transcendental ($I_4$, $I_5$) functions,
	the eigensystem solution cannot be calculated in the typical way.
	As a {\it first step} to solve the complete eigenvalue problem, the transcendental part is discarded and,
	\eqref{eq:Dxwx} is multiplied by the lowest common denominator obtaining thus a polynomial matrix.
	Such a matrix can be written as a polynomial in $\omega$ with constant matrix coefficients $M_n$.
	This polynomial is reorganized to obtain the following matrix pencil \cite{Markus2012,ManfredMoeller2015}
	\begin{subequations}
	\begin{equation}
		\mathcal{D}^\#(\omega)\vecx = M_0\vecx_0 + M_1\vecx_1 + \cdots + M_n\vecx_n - \omega M_{n+1}\vecx_n =0,
	\end{equation}
	where
	\begin{equation}
		\vecx_n = \omega^n\vecx.
	\end{equation}
	\end{subequations}
	
	\begin{figure}[t]\centering
		\includegraphics[width=0.5\textwidth]{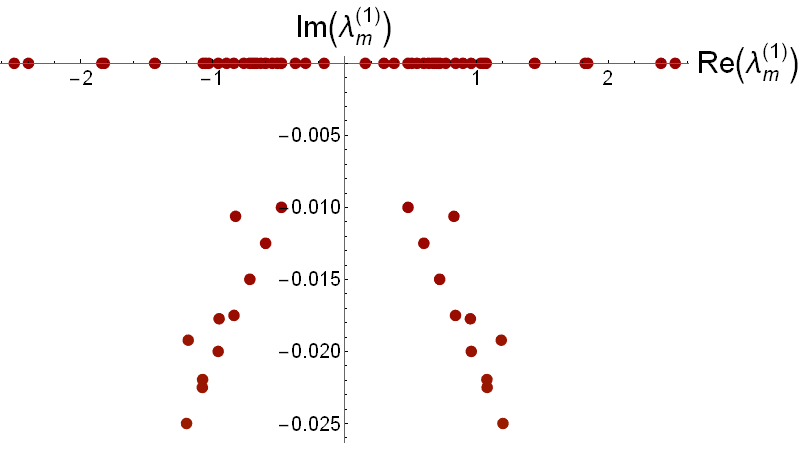}
			\caption{Eigenvalues $\lambda_m^{(1)}$ of the algebraic part of $\mathcal{D}(\omega)$,
			using the values in table \ref{tab:tbc} with 12 modes.}
		\label{fig:MDSN7-eV-C13-M12}
	\end{figure}%
	
	Subsequently, the matrix pencil is reorganized in a higher dimension as
	\begin{align}\resizebox{0.96\hsize}{!}{$
		\left(\begin{array}{@{}c|ccc@{}} M_0 & M_1 & \cdots & M_n \\ \hline \\[-10pt]
			0 \\ \vdots & & \mathbb{I}_n\otimes\mathbb{I} & \\[4pt] 0 \end{array}\right)
		\begin{pmatrix} \vecx_0 \\ \vecx_1 \\ \vdots \\ \vecx_n \end{pmatrix} = \omega
		\left(\begin{array}{@{}ccc|c@{}} 0 & \cdots & 0 & M_{n+1} \\ \hline & & & 0 \\ & \mathbb{I}_n\otimes\mathbb{I} & & \vdots \\[4pt] & & & 0 \end{array}\right)
		\begin{pmatrix} \vecx_0 \\ \vecx_1 \\ \vdots \\ \vecx_n \end{pmatrix},
	$}\end{align}
	and by multiplying the equation by the inverse of the matrix on the right side,
	the equation is linearized as
	\begin{equation}\label{eq:Dwlin}\resizebox{0.92\hsize}{!}{$
		\left(\begin{array}{@{}c|ccc@{}} 0 \\ \vdots & & \mathbb{I}_n\otimes\mathbb{I} & \\[4pt] 0 \\ \hline \\[-10pt]
			M_{n+1}^{-1}M_0 & M_{n+1}^{-1}M_1 & \cdots & M_{n+1}^{-1}M_n \end{array}\right)
		\begin{pmatrix} \vecx_0 \\ \vecx_1 \\ \vdots \\ \vecx_n \end{pmatrix} =
		\omega \begin{pmatrix} \vecx_0 \\ \vecx_1 \\ \vdots \\ \vecx_n \end{pmatrix},
	$}\end{equation}%
	so the eigenvalues $\lambda_m^{(1)}$ of the algebraic part are trivially found.
	Although it is necessary to take into account all the $\lambda_m^{(1)}$, in the case of a tandem configuration,
	as in any other case in the computation of Green's functions,
	it is imperative to eliminate those with $\Im[\lambda_m^{(1)}] >0$
	for convergence of the integrals that include the resolvent of the wave operator,
	avoiding thus the non-physical gains, so the integration path in the definition of
	the Green's function is the usual that encapsulates the poles in the lower plane \cite{MoshinskySadurni2007}.
	The resulting pole structure is shown in fig.~\ref{fig:MDSN7-eV-C13-M12},
	that has the property $-\omega = \omega^*$ in order to maintain causality.
	
	Since the Euclidean norm of the algebraic part of (the finite dimensional) $\mathcal{D}(\omega)$
	in \eqref{eq:Dxwx} is greater or similar to that of its transcendental part,
	the eigenvalues of the complete problem $\lambda_m^{(2)}\simeq\lambda_m^{(1)}$.	
	Therefore, as a {second step}, we take a small variation by a parameter $\xi$ in each of the $\lambda_m^{(1)}$ found above,
	so we now have to solve for an operator $\mathcal{D}(\omega)$ that considers the transcendental part, i.e.
	\begin{equation}
		\mathcal{D}(\lambda_m^{(1)}+\xi)\vecx = \frac{(\lambda_m^{(1)}+\xi)^2}{c^2}\vecx.
	\end{equation}
	The above expression is expanded in $\xi$ to low order,
	and the construction of the matrix pencil in $\xi$ for each $\lambda_m^{(1)}$ gives the correction
	$\lambda_m^{(2)} = \lambda_m^{(1)} + \xi_m$.
	This process is performed iteratively until the desired precision is obtained, i.e.
	\begin{equation}
		\det\left( \mathcal{D}(\lambda_m) - \frac{\lambda_m^2}{c^2 }\mathbb{I}\right) \rightarrow 0,
	\end{equation}
	so the frequencies $\omega_m$ are encountered.
	Finally, the null space of \eqref{eq:Dxwx} is calculated for each $\omega_m$, obtaining $v_n(\omega_m)$.
	This method is arbitrarily precise. If the logarithm is not approximated it is still possible to use a matrix pencil,
	but the matrix in \eqref{eq:Dwlin} will contain terms with logarithms.

%

\end{document}